\documentclass[fleqn,usenatbib]{mnras}

\usepackage{newtxtext,newtxmath}
\usepackage{xspace}

\usepackage{natbib}  
\setcitestyle{round,aysep={},yysep={;}}  

\usepackage{graphicx}	
\usepackage{amsmath}	
\usepackage{float}

 \title[Age of high-z JWST galaxies]{Improved measurements of the age of JWST galaxies at $z=6-10$}
\author[L\'opez-Corredoira \& Guti\'errez]{M. L\'opez-Corredoira$^{1,2}$\thanks{E-mail: martin@lopez-corredoira.com}, C. M. Guti\'errez$^{1,2}$
\\
$^1$ Instituto de Astrof\'\i sica de Canarias, E-38205 La Laguna, Tenerife, Spain\\
$^2$ Departamento de Astrof\'\i sica, Universidad de La Laguna,
E-38206 La Laguna, Tenerife, Spain
}

\date{Last Rev. 23 January 2026}

\begin{document}
\label{firstpage}
\maketitle

\begin{abstract}
From {\it James Webb Space Telescope (JWST)} surveys, 31 galaxies with average redshift 7.3 are selected containing large Balmer break, Lyman-$\alpha$ break (V-shaped SED versus $\lambda$). Apart from {\it Hubble Space Telescope (HST)} and {\it JWST}-NIRCam (Near-infrared camera) photometry for these galaxies, there are {\it JWST}-NIRSpec (Near-infrared spectrograph) spectra for 13 galaxies and mid-infrared photometry (mostly {\it JWST}-MIRI) for 15 of them. Spectroscopical analyses included Balmer emission lines, Balmer + 4000 \AA \ breaks or CaII lines. Spectral energy distribution (SED) fitting with photometry include old and young stellar populations, emission lines associated to HII regions, AGN, interstellar dust extinction and intergalactic extinction from neutral hydrogen.
By adopting realistic extinction curves and taking into account the V-shaped SED and low emission at near infrared at rest, the analyses show that AGN contribution in these galaxies ('little red dots' most of them) should be small on average in the reddest wavelengths, though important for few of the 31 galaxies.
Average age of the 31 galaxies: $0.61\pm 0.31$(95\% CL) Gyr, while the average age of the $\Lambda $CDM universe is 0.70 Gyr. This corresponds to a formation epoch $z_{\rm form.}>11.2$(97.5\% CL).
Reddest galaxies present largest ages. One of these very red galaxies gets an age incompatible to be younger than the age of the Universe within $>4.7\sigma$. TP-AGB effect cannot explain this tension. None the less, there may be other uncertainties in the models, so this tension is a provisional result and further research is needed to confirm it.
\end{abstract}


   \maketitle
%

\section{Introduction}
\label{.intro}

\citet{Lab23} identified 13 candidate massive galaxies ($\gtrsim 10^{10}$ M$_\odot $
stellar mass on average) with average $z\sim 8$ (age
of the Universe with the standard cosmological model $\sim 600$ Myr), a result that is at odd with knowledge on the formation of galaxies. They were selected photometrically from
{\it HST (Hubble Space Telescope)} + {\it JWST(James Webb Space Telescope)}-CEERS (The Cosmic Evolution Early Release Science Survey), requiring the presence of large Balmer break, Lyman-$\alpha $ break and suppressed emission at shorter wavelengths of far-ultraviolet (UV) at rest corresponding to a lack of detection with {\it HST} visible bands.

\citet{Lop24} (hereafter Paper I) estimated the average age of the oldest stellar population
in these 13 galaxies to be on average larger than 900 Myr within 95\% CL, a result which is in tension with the age of the Universe in the standard model; similar tension was also obtained by \citet{Mar25} with similar data and technique.
 For this estimation in Paper I, a fit of Spectral energy distribution (SED) obtained from photometry in 10 filters
at visible and near infrared was carried out assuming a combination of two Single stellar populations (SSPs) with the same
extinction; a lower extinction for the younger population would lack physical justification, and considering the 
presence of the Ly-$\alpha $ break this extinction should be low, avoiding degeneracy age-dust at longer
wavelengths \citep{Lop17,Lop24,Gao24}.
Extra stellar populations did not significantly improve the fits. Also, 
the inclusion of an active galactic nucleus (AGN) component was investigated by Paper I(Sect. 3.4),
assuming an independent AGN extinction with $A_V\le 3$ magnitudes and it was concluded that,
although a minor component due to AGN is possible on the average SED, its inclusion in the fit 
does not significantly change the range of
the age of the oldest stellar population necessary to produce the fit. 
However, some possibilities of interstellar extinction 
within $A_V\le 3$ and the effect of the intergalactic extinction at reionization epoch was not explored.

When \citet{Lab23} published their work based on their selection of 13 galaxies candidates to $z\gtrsim 6$ massive galaxies, only the photometric fluxes were available
and none of the galaxies had been observed spectroscopically. Now spectra of 6 of these 13 galaxies have been obtained with {\it JWST}-NIRSpec (Near-infrared spectrograph; 1-5 $\mu$m),
\footnote{https://dawn-cph.github.io/dja/spectroscopy/nirspec/ } with spectroscopic redshifts approximately confirming their high photometric redshifts, and providing some extra clues on the nature of these objects. So far, two of them present
broad emission lines: \#C-13050 [here the same galaxy number than \cite{Lab23} is used, Paper I] 
with FWHM-H$_\beta $=1800 km/s \citep{Koc23}; and \#C-38094 with FWHM-H$_\beta $=3600 km/s \citep{Wan24} (hereafter W24), indicating
the presence of at least some partial component of AGN. The galaxy \#13050 was limited to have a black hole mass few times $10^7$ solar masses, possibly 
negligible for its continuum flux contribution, but
the \#C-38094 would have black hole mass larger than 10$^8$ M$_\odot $ (and even larger than 10$^9$ M$_\odot $ 
if the extinction apparently derived
in some SED fitting is assumed), 
which makes us think that the presence of an AGN component among the 13 selected massive galaxies is not
negligible, as also pointed out by other analyses \citep{Bar24,Chw24,Li25}.
Moreover, as it is also obtained by the spectral fits carried out by W24,
extinction reaching values as high as $A_V>3$ for the AGN cannot be excluded.

In a similar survey with similar characteristics' galaxies, \citet{Per24} obtained that 3 (none of them within the selection in Table \ref{Tab:bestfits}) out of 18 red galaxies
with available spectra present broad line features. It seems that AGN-Seyfert 1 may be a component of a small part of the sample of this kind of galaxies, but most of them lack any sign of it.
\citet{Per24} also claimed that they are less massive in stellar component than
the numbers obtained by \citet{Lab23}. However,
there are also other massive or ultramassive galaxies at these redshifts
\citep[e.g.,][]{Xia24}.

W24 carried out a fit over the spectra of the two galaxies common with \citet{Lab23} sample, 
and concluded that the most likely fit for both of them contains red-QSO components that are dominant at $\lambda \gtrsim 4800$ \AA $_{\rm rest}$, whereas for $\lambda \lesssim 4800$ \AA $_{\rm rest}$ non-AGN stellar population dominate with ages of the oldest stellar population of 
500 - 600 Myr. The presence of pronounced Balmer breaks make necessary
the old stellar component \citep{Set24,Ma25} [see however another proposal in terms of AGNS surrounded by clumps of extremely
dense gas \citep{Ina25}], 
whereas the broad Balmer lines at least for one of these galaxies (\#C-38094) may make necessary the AGN component.
However, some models suggest that pure stellar populations could also explain the observations. In this case, the broad Balmer lines might result from extremely fast stellar rotation in ultra-compact galaxies, from star-formation-driven outflows in dense or low-metallicity environments, or from inelastic Raman scattering of stellar UV light by neutral hydrogen (W24)\citep{Bag24,Kok24}. Alternatively, scenarios with even higher AGN contributions have been proposed, though they struggle to account for the observed Balmer break.
Indeed,  with the combination of red-QSO and stellar population that best fits the data, 
the Balmer breaks of the theoretical model are
much smaller than the observed ones (W24, fig. B1), indicating the need of still older stellar populations in case an important AGN is included.
 
As W24 or \citet{Ma25} recognize, their SED fitting with unexplained features and their interpretation present some gaps in an incomplete understanding of these galaxies. Moreover, 
some parameter values in their best fits appear inconsistent with expectations: they get very massive black holes, in consonance with the trend to find JWST high-z ``little red dots'' (LRDs), mysterious objects that in some sense resemble massive black holes with huge reddening.
\citet{Lin25} suggested that compact dwarf galaxies at low redshift are similar to LRDs, 
hosting broad-line AGNs and other various similar properties, thus positing they are the same class of object with different evolution stage. However, this hypothesis clashes with the common understanding
of the formation of black hole in the early Universe and moreover lack of typical signs of AGN such as UV variability or X-ray emission
\citep{Kok24} or radio emission \citep{Per25}.
The lack of X-ray emission might be due to the central black holes accrete
at super-Eddington rates, where X-ray emission may be
intrinsically weak \citep{Pac24} or it 
may be absorbed by dust-free, high column
density gas in the broad-line region \citep{Mai24}, but still it is 
challenging to reconcile the number of coincidences under standard assumptions of massive black holes.
Furthermore, the extinction curve W24 obtain for their best fits for the AGN component is very different from the properties of dust usually
admitted (see Sect. \ref{.ext}).
Another possible interpretation of these objects is to be extreme starburst star clusters precursors of present-day massive globular clusters \citep{Jer17}.

W24 and Paper I interpretations deserve further analyses. In the first case, W24 do not get a satisfactory physical interpretation of their mathematical fits, and, like the vast majority of researchers doing
SED fitting of high-$z$ galaxies, have set a constraint in their models to avoid the stellar populations being older than the Universe. 
This is an aspect that requires careful testing before being accepted as an established fact. Some studies show that better fits are obtained when allowing for galaxy ages older than the age of the Universe (Paper I, \citet{Mar25,Ste24}). Moreover, since one of our aims is to use these observations as a cosmological test, it is essential to avoid adopting any specific assumptions a priori.
 In the second case, authored by one of us, Paper I has not taken into account the possibility
of AGN with extinctions $A_V>3$ and intergalactic extinction.
Furthermore, the information on 
the emission lines, obtained by the spectra now available, and few available photometric data of some of these galaxies 
at 5 $\mu$m$<\lambda \le 15\ \mu$m and a larger number of galaxies can be considered now. 
These new informations or other minor issues are included here. This is carried out in the following sections.
Sect. \ref{.data} introduces the photometric {\it HST}+{\it JWST} (visible, near-infrared and mid-infrared)  data and near-infrared JWST spectra to be used in this paper. Sect. \ref{.ext} presents the extinction curve to be used, with special attention to the absorption below 1\,000 \AA \ at rest. Spectra are analysed in Sect. \ref{.spectra}, whereas SED fits of the photometry are carried out in Sect. \ref{.phot}. Summary, discussion and conclusions are given in Sect. \ref{.disc}.

\section{Data}
\label{.data}

Two sets of data are used:
\begin{itemize}
\item The same 13 galaxies analyzed by \citet{Lab23} and Paper I, from {\it HST}+{\it JWST}-CEERS \citet{Fin23}, whose fluxes at 10 filters with $\lambda <5\ \mu$m are 
given at \citet[Extended Data Table 1]{Lab23}. The selection criteria were:
F150W-F277W$<$0.7, F277W-F444W$>$1.0 (V-shaped), with good SNR (signa-noise ratio) [F444W$<27$ AB; F150W$<29$ AB; SNR(F444W)$>8$].
These criteria were chosen by \citet{Lab23}: (i) to ensure no secondary redshift solutions at low-z, and (ii) to filter massive galaxies with high Mass/Luminosity ratios. The (photometric or spectroscopic) redshift of these sources is between 5.6 and 9.8 (see Table \ref{Tab:bestfits}).
Galaxies were selected to have no nearby companions.
They are at northern equatorial coordinates. ID indicated
as C-xxxxx. 

\item New 18 galaxies selected from the 
photometric survey {\it HST}+{\it JWST}-JADES (The JWST Advanced Deep Extragalactic Survey) \citep{Eis23,Rie23,Bun24}, 
at 22
filters $\lambda <5\ \mu$m,
with the same color and SNR criteria of \citet{Lab23}, in the fields GOODS-S (Great Observatories Origins Deep Survey-South) (J-0xxxxxx) and GOODS-N (Great Observatories Origins Deep Survey-North) 
(J-1xxxxxx). It is the complete sample within these criteria. The motivation of the selection criteria is the same one than the previous point.
Galaxies were selected using quality flags to ensure reliability and to have no nearby companions. The (photometric or spectroscopic) redshift of these sources is between 5.7 and 8.4 (see Table \ref{Tab:bestfits}).
\end{itemize}

New fluxes are also added, of 
observed 5 $\mu$m$<\lambda \le 26\ \mu$m fluxes from IRAC \citep{Faz04} or {\it JWST}-MIRI (Mid-infrared Instrument)\citep{Wri23}, provided in Table \ref{Tab:newfluxes}.
The data from IRAC (Infrared Array Camera) were obtained under crosscorrelation of sources with a limiting
maximum distance of 1.2 arcsec, the IRAC pixel size, which is around 3 times the precission of IRAC
astrometry [around 0.37 arcsec \citep{Bon19}].

\begin{table*}
\caption{Available fluxes at 5 $\mu$m$<\lambda \le 26\ \mu$m for some of the 31 galaxies used here.
Fluxes' units are nJy. 
Sources: {\it Spitzer}-IRAC [5.8, 8.0 $\mu$m] \citep{Guo13,Ste17,Oes18,Ste21}; 
{\it JWST}-MIRI [5.6, 7,7, 10.0, 12.8, 15.0, 18.0, 25.5 $\mu$m] (\citet{Bar24,Rie24};
https://archive.stsci.edu/hlsp/smiles ).}
\begin{center}
\begin{tabular}{cccccccccc}
Galaxy ID & F$_{\nu ,5.6\mu m}$ & F$_{\nu ,5.8\mu m}$ & F$_{\nu ,7.7\mu m}$ &
F$_{\nu ,8.0\mu m}$ & F$_{\nu ,10.0\mu m}$ &
F$_{\nu ,12.8\mu m}$ & F$_{\nu ,15.0\mu m}$ 
& F$_{\nu ,18.0\mu m}$ & F$_{\nu ,25.5\mu m}$ \\ \hline
C-2859 &  --& -- & -- & -- &  700$\pm$100 & -- &  $350\pm 350$ & -- & -- \\
C-7274 & -- & 70$\pm $1220 & -- & 360$\pm $910  & -- & --  & -- & -- & -- \\
C-11184 & -- & -470$\pm $970 & -- & 840$\pm $770  & -- & -- & --  & -- & -- \\
C-14924 & -- & -290$\pm $1010 & -- & 220$\pm $900  & -- & --  & -- & -- & -- \\
C-38094 & -- & 2010$\pm $1840 & -- & 3050$\pm $1590  & -- & -- & -- & -- & -- \\ \hline
J-0066293 & 365$\pm $52 & -- & 212$\pm $42 & -- & 199$\pm $93 & 185$\pm $157 & 17$\pm $197 & -- & --\\  
J-0075634 & 388$\pm $54 & -- & 225$\pm $43 & -- & 95$\pm $95 & 139$\pm $173 & 274$\pm $201 & 362$\pm $533 & --\\ 
J-0165305 & 276$\pm $69 & -- & 31$\pm $55 & -- & 176$\pm $119 & -- & -- & 318$\pm $791 & 4409$\pm $8230 \\   
J-0204022 & 319$\pm $45 & 355$\pm $414 & 206$\pm $44 & 1090$\pm $480 & 181$\pm $93 & -- & 215$\pm $207 & 1216$\pm $546 & 8255$\pm $ 6439  \\
J-0210600 & 77$\pm $35 & 210$\pm $198 & 178$\pm $30 & 277$\pm $257 & 237$\pm $72 & 249$\pm $137  & 301$\pm $178 
& 763$\pm $478 & -- \\
J-0211388 & 119$\pm $53 & -- & 118$\pm $42 & -- & 410$\pm $93 & 651$\pm $155 & 367$\pm $199 & 1202$\pm $518 & 629$\pm $5834 \\
J-0214552 & 193$\pm $52 & -- & 296$\pm $47 & -- & 283$\pm $93 & 155$\pm $154 & 117$\pm $202 & 
364$\pm $528  & -- \\ \hline
J-1010816 & -- & 143$\pm $156 & -- & 233$\pm $173 & -- & -- & --  & -- & -- \\
J-1032447 & -- & -- & -- & 127$\pm $270 & -- & -- & -- & -- & -- \\
J-1050323 & -- & 693$\pm $217 & -- &803$\pm $245 & -- & -- & -- & -- & -- \\
\hline
\end{tabular}
\end{center}
\label{Tab:newfluxes}
\end{table*}

The photometric points will be used for SED fitting in Sect. \ref{.phot}.
Thirteen of these 31 galaxies have {\it JWST}-NIRSpec spectra (1-5 $\mu$m), and this will be used for a separate spectral analysis
in Sect. \ref{.spectra}. Spectral resolution is low ($R=$100; prism-clear) except for C-39575 and J-210600 in which medium resolution
spectra ($R=$1000) are used. J-* galaxies are obtained and reduced within the JADES ({\it JWST} Advanced Deep Extragalactic Survey) programme. Origin of the spectra in CEERS galaxies is indicated
in Table \ref{Tab:bestfits}. 

Although the angular size of these galaxies was not used as a selection criterion, 18 of our 31 galaxies are included in the 
catalog of LRDs by \citet{Per25}; possibly the ratio might be higher given that the sample of LRDs by \citet{Per25} is incomplete.
Therefore, most of our sources are LRDs.

\section{Extinction curves}
\label{.ext}

\subsection{Interstellar extinction}

\begin{figure}
\vspace{0cm}
\centering
\includegraphics[width=8cm]{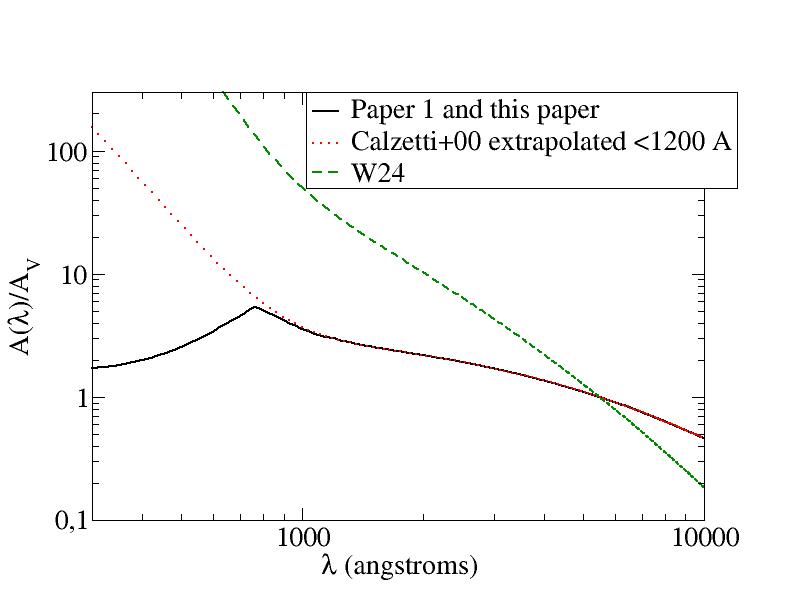}
\vspace{.2cm}
\caption{Log-log plot of interstellar extinction curves used in W24, Paper I and this paper. 
All of them are based on \citet{Cal00} curve (only valid for 
$\lambda \ge 1200 $ \AA ), but for Paper I and herein \citet{Wei01}-$R_V=4$ curve for $\lambda _{\rm rest}<1200$ \AA \ is used,
whereas for W24 \citet{Cal00} curve is extrapolated for $\lambda _{\rm rest}<1200$ \AA \ and added a factor 
$\left(\frac{\lambda }{5500\ \AA}\right)^{-1.53}$.}
\label{Fig:ext}
\end{figure}

Figure \ref{Fig:ext} shows the extinction curves used in Paper I and W24.
Paper I used \citet{Cal00} curve for $\lambda _{\rm rest}\ge 1200 $ \AA \ and \citet{Wei01}-$R_V=4$ curve for $\lambda _{\rm rest}<1200$ \AA . The main motivation to adopt this extinction curve is that \citet{Cal00} is only valid for $\lambda>1200$ \AA, 
and an extrapolation of their law does not reproduce some features at lower wavelengths, particularly the bump around 800 \AA. This is clearly observed in Fig. \ref{Fig:ext}.
However, W24 used \citet{Cal00} curve for the
whole wavelength range with a flexible power-law slope proposed by \citet{Nol09}: $A(\lambda )=A_{\rm Calzetti}(\lambda) 
\left(\frac{\lambda }{5500\ \AA}\right)^{n_{\rm dust}}$, with average $n_{\rm dust,AGN}=-1.53$ for galaxies 
\#C-13050 and \#C-38094. Similar assumptions were adopted by \citet{Ma25}. 
There are other proposals of extinction curve in the literature \citep[e.g.,][Fig. 1]{Li25},
but they do not affect the order of magnitude of the extinction at $\lambda _{\rm rest}>1000$ \AA .

As can be observed
in Fig. \ref{Fig:ext}, the interstellar extinction used for the best fits by W24 gives extremely large values of extinction in far-UV-rest wavelengths, which are very far from the realistic modelling by \citet{Wei01} and \citet{Gor09} to reproduce extinction in galaxies of the local Universe. The QSO reddening also follows this extinction curve, typical of
a Small Magellanic Cloud (SMC) attenuation law, which is assumed to be a good description for the reddening in Type 1 SDSS (Sloan Digital Sky Survey) QSO spectra \citep{Cal21}. For this attenuation law, 
it is well known that far-UV wavelenghts, $\lambda \lesssim 800$ \AA , get a strong decrease of extinction, allowing the far-UV light to pass through thick layers of dust, 
whereas near-UV is more opaque.
Furthermore, W24 uses $n_{\rm dust,AGN}=-1.53$, which is also far from any realistic modelling of dust
in the conditions that are known in the local Universe galaxies. \citet{Nol09} proposed their modification of \citet{Cal00} curve by
a factor $\left(\frac{\lambda }{5500\ \AA}\right)^{n_{\rm dust}}$ to correct small deviations, but only allowing $|n_{\rm dust}|<0.3$ in 
realistic cases. A value of $n_{\rm dust}\approx -1.5$ is very far from a realistic description of dust. 
However, as we will see in the next section, the large intergalactic extinction can compensate in an order of magnitude the deficit of interstellar extinction.

In addition, W24 allows a variable fraction (another free parameter) 
of starlight outside the dust screen, permitting very young blue stars to be without extinction.
This is a new artefact useful to produce better fits mathematically speaking, but without
a realistic physical interpretation (Paper I): blue stars should have associated dust since
they are embedded in star formation regions. And extinction $A_\lambda $ should represent an
average extinction of the whole population, for which holes simply reduce the average but
they are not selectively applied over blue stars.

The above reasons have motivated us to keep a standard
extinction curve modelled by \citet{Cal00} for $\lambda _{\rm rest}\ge 1200 $ \AA  \
and \citet{Wei01}-$R_V=4$ curve for $\lambda _{\rm rest}<1200$ \AA , like in Paper I.

\subsection{Intergalactic extinction}

\begin{figure}
\vspace{0cm}
\centering
\includegraphics[width=8cm]{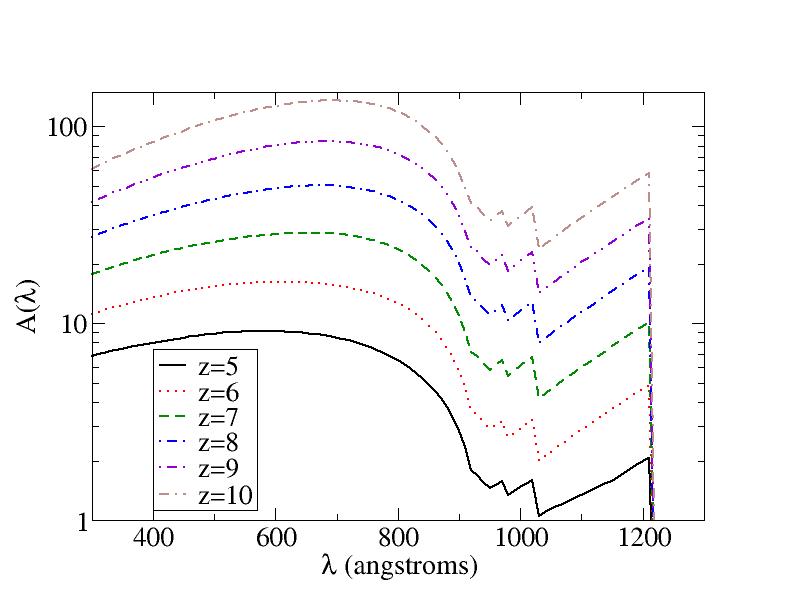}
\vspace{.2cm}
\caption{Log-linear plot of intergalactic extinction curves derived by \citet[sect. 4]{Ino14}.}
\label{Fig:extIGM}
\end{figure}

Attenuation of the radiation from distant objects by intergalactic neutral hydrogen should also be considered for $\lambda _{\rm rest}<1216$ \AA \
in galaxies at high-$z$, where reionization epoch takes place. 
Here we adopt the analytical expressions given by \citet[sect. 4]{Ino14}. This considers the contributions of Lyman series and Lyman continuum
for the Ly$\alpha $ forest, Lyman-limit systems and damped Ly$\alpha $ systems.
Figure \ref{Fig:extIGM} shows the attenuation $A(\lambda )$ for different redshifts.

\section{Spectral analysis}
\label{.spectra}

\begin{figure}
\vspace{0cm}
\centering
\includegraphics[width=8cm]{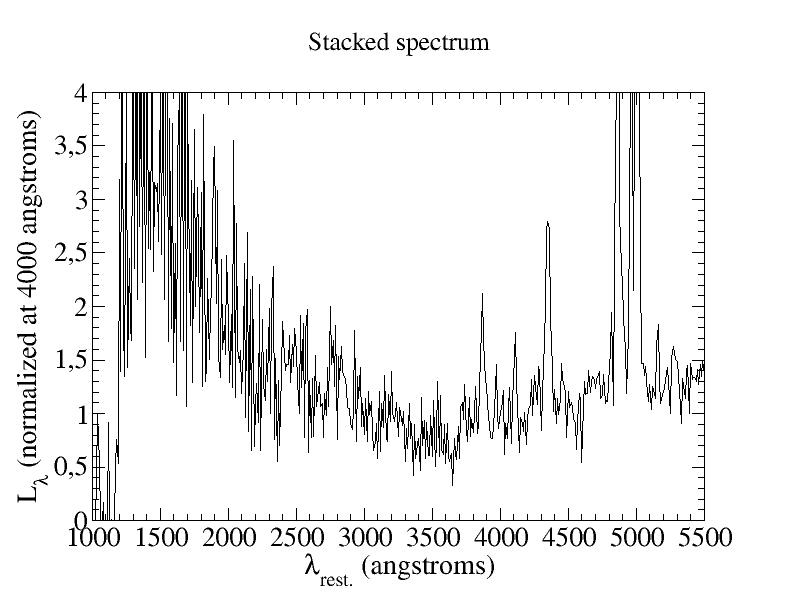}
\vspace{.2cm}
\includegraphics[width=8cm]{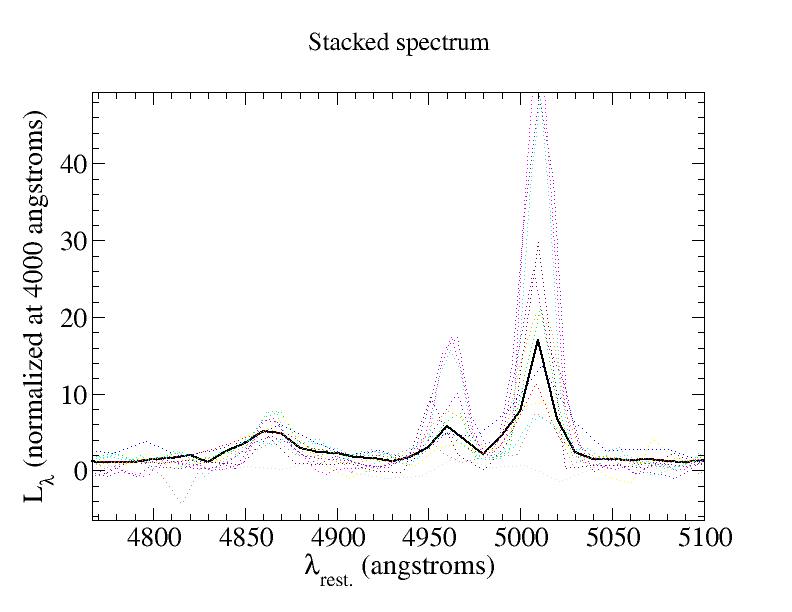}
\vspace{.2cm}
\caption{Top panel: Stacked spectrum (black solid line) corresponding to the weighted average of the 13 galaxies (normalized
each of them at 4000 \AA ) with available spectrum, in the range $\lambda _{\rm rest}=$1000-5500 \AA \ at rest. 
Resolution $\Delta \lambda =10$ \AA . Bottom panel: 
Zoom of the top panel in the range $\lambda _{\rm rest}$=4770-5100 \AA ; dotted lines stand for the individual 13 spectra.}
\label{Fig:suma}
\end{figure}

{\it JWST}-NIRSpec spectra (1-5 $\mu $m) are available for the 13 galaxies with spectroscopic redshift (see Table \ref{Tab:bestfits};
corresponding to the 13 galaxies with redshifts without errors).
Fig. \ref{Fig:suma} shows the stacked normalized spectra of these 13 galaxies in the range
1000 \AA $<\lambda _{\rm rest.}<$5500 \AA , common to all of them.
The stacking corresponds to the weighted average of the 13 galaxies (normalized
each of them at 4000 \AA ) with available spectrum, with $\Delta \lambda =10$ \AA .

In Table \ref{Tab:spectra}, some measurements are given, corresponding to the features stacked spectra 
including all of them or for different colours F277W-F444W. The separation in colour is justified given the
correlation colour-age (see Sect. \ref{.color}).
They correspond to:

\subsection{H$_\beta $, H$_\gamma $ and its corresponding $A_V$} 
\label{.hbeta}

An intrinsic (without extinction) ratio $\frac{F_\lambda(H_\gamma)}{F_\lambda(H_\beta)}=0.46$ \citep[ch. 13]{Ost06} is 
assumed.
The error bars are too large for an accurate determination of the extinction. 
The extinction might be derived instead from H$_{\alpha}$/H$_{\beta }$ for few galaxies (only the galaxies 
with $z<6.9$), but this is subject to more systematic
errors \citep{Ost06,Ill12,Wu23} in deriving the amount of extinction given the sensitivity of the ratio
to other parameters different from reddening.

\subsection{Balmer and 4\,000 \AA \ breaks} 

There are many different definitions of 4\,000 \AA \ break, 
including ranges over and below 4,000 \AA : either wide range D4000 \citep[e.g.,][]{Bin19,Ste24,Wan24,Rob24} or narrow range Dn4000 \citep[e.g.,][]{Bal99,Lop18}. Use of narrow ranges around 4\,000 \AA \ has the advantage of being less affected by extinction, but one should avoid some emission lines like NeIII-3869 and H$_\delta $-4102.
 
\begin{figure}
\vspace{0cm}
\centering
\includegraphics[width=8cm]{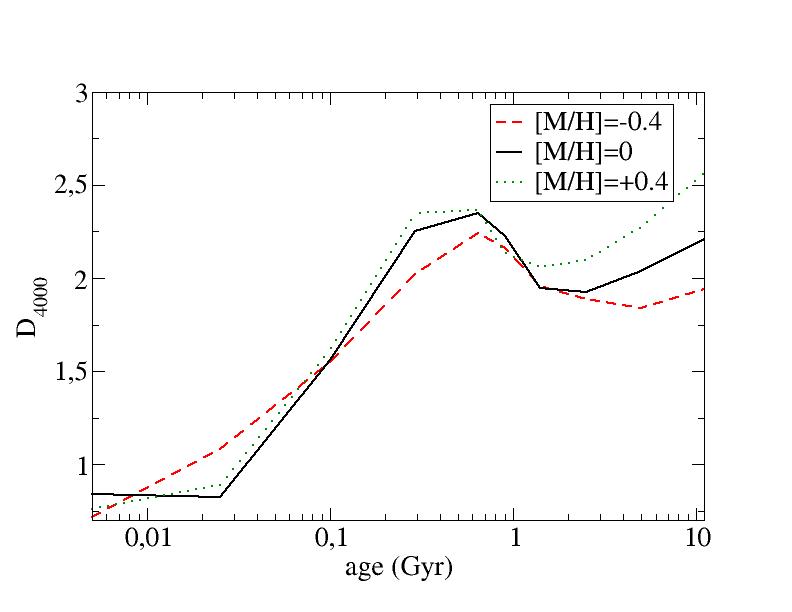}
\vspace{.2cm}
\includegraphics[width=8cm]{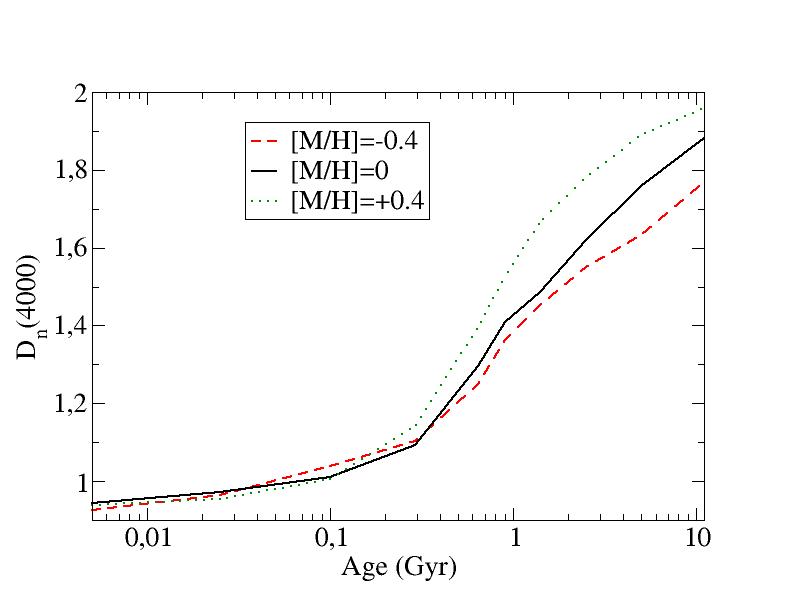}
\vspace{.2cm}
\caption{Prediction of $D(4000)=\frac{F_\lambda(4160-4290)}{F_\lambda(3500-3630)}$ 
and $D_n(4000)=\frac{F_\lambda(4022-4072)}{F_\lambda(3900-3950)}$ for a single
stellar population in GALAXEV stellar population synthesis model, as a function of age and metallicity.}
\label{Fig:D4000}
\end{figure}

Here two definitions are adopted, wide and narrow respectively, avoiding the spectrum ranges with some emission lines:

\begin{description}
\item[Balmer break indicator $D(4000)$]
\begin{equation}
D(4000)=\frac{F_\lambda(4160-4290)}{F_\lambda(3500-3630)}
\end{equation}
 used by \citet{Rob24} and \citet{Wit24} [note however that these authors use the definition of the break parameter $B$ with $F_\nu $ instead of $F_\lambda $; hence $B=1.405\,D(4000)$], including
the Balmer break. For the stacked spectrum $D(4000)=1.43\pm 0.13$ ($B=2.01\pm 0.18$ in $F_\nu $) is obtained,
and for the stacking of the four galaxies with F277W-F444W$>$1.7 $D(4000)=1.74\pm 0.17$ ($B=2.44\pm 0.24$ in $F_\nu $)
is obtained.

Breaks in Balmer (3645 \AA ) and/or 4000 \AA \ are related to the age of the 'average' stellar populations, and with a minor dependence on metallicity too. In order to derive an estimate of the 'oldest' population age, subtraction of the component due to the non-old component is needed \citep{Lop18}. An extinction correction should also be applied. 
If only two SSPs and AGN components are 
assumed, with $r_2=\frac{F_{\lambda ,{\rm young}}(3565\ \AA )}{F_{\lambda ,{\rm old}}(3565\ \AA )}$,
$r_3=\frac{F_{\lambda ,{\rm AGN}}(3565\ \AA )}{F_{\lambda ,{\rm old}}(3565\ \AA )}$,
interstellar extinction $A_V$ at V-magnitude,  $A_{3565\ \AA}-A_{4225\ \AA}=0.20\,A_V$,
AGN extinction $A_{V,AGN}$,
\begin{equation}
D_{\rm old}(4000)=-r_2\,D_{\rm young}(4000)
\end{equation}\[
-r_3\,D_{\rm AGN}(4000)\times 10^{0.4\times 
0.20\times (A_{V,AGN}-A_V)}
\]\[
+D(4000)\times (1+r_2+r_3)\times 10^{-0.4\times 0.20\times A_V}
,\]
where 
$D_{\rm old}(4000)\equiv D_{\rm 1SSP}(4000)[age_{\rm old},[M/H]]$, 
$D_{\rm young}(4000)\equiv D_{\rm 1SSP}(4000)[age_{\rm young},[M/H]]$, 
and $D_{1SSP}$ is given in Fig. \ref{Fig:D4000}/top panel using the same model 
GALAXEV \citep{Bru03} that is used for SED fits at
Sect. \ref{.results}. Assuming age$_{\rm young}$=10 Myr, [M/H]=0 (the
10 Myr age assumed is based on the best-fitting photometric models, which consistently
yield this value; see Table \ref{Tab:bestfits}), $D_{\rm young}(4000)$=
0.83, almost independent of metallicity.
The corresponding break for the AGN template is $D_{\rm AGN}(4000)=0.60$.

If we use the average values of the parameters $r_2=0.15^{+0.32}_{-0.08}$, $r_3=0^{+0.88}{-0}$ and extinctions
$A_V=0^{+0.75}_{-0}$, $A_{V,AGN}=0^{+\infty}_{-0}$, 
as obtained from the best SED fit of the average galaxy
[Table \ref{Tab:bestfits}/[SAB], Fig. \ref{Fig:sedrstacked}/upper panel);
$r_2\approx 5\times A_2\times 10^{-0.4\times 
0.50A_V}$, $r_3\approx 7\times A_3\times 10^{-0.4\times 0.50(A_{V,AGN}-A_V}$)] 
we get $D_{\rm old}(4000)=1.64^{+0.77}_{-0.26}$ (68\% CL), and,
through the relationship of Fig. \ref{Fig:D4000}/top panel,
$age_{\rm old}>60$(68\% CL),$>30$ Myr (95\% CL) Myr 
for the stacking of the 13 galaxies. 
  
For the stacking of the four galaxies with F277W-F444W$>$1.7,
$D_{\rm old}(4000)=1.88^{+1.06}_{-0.34}$ (68\% CL), $age_{\rm old}>90$ Myr (68\% CL), 
$>40$ Myr (95\% CL).
Note in any case that these ones are just rough estimations based on the hypothesis of the existence
of two dominant SSPs and an AGN and assuming values of the parameters 
derived from the SED fits in the next section. The true errors might be larger.


\item[Narrow Balmer break indicator $D_n(4000)$]
\begin{equation}
D_n(4000)=\frac{F_\lambda(4022-4072)}{F_\lambda(3950-3900)}
\end{equation}
similar to \citet{Lop18} but shifted -28 \AA \ in the longer wavelengths to keep a distance larger than 30 \AA \ 
from H$_\delta $-4102 line.
For the stacked spectrum,
$D_n(4000)=1.08\pm 0.08$; and for the stacking of the four galaxies with F277W-F444W$>$1.7, 
$D_n(4000)=1.14\pm 0.10$.

 Similarly of the above procedure, the contribution of the old population can be estimated
for the narrow break indicator: 
\begin{equation}
D_{n,\rm old}(4000)=-r_{2n}\,D_{n,\rm young}(4000)
\end{equation}\[
-r_{3n}\,D_{\rm n,AGN}(4000)+D_n(4000)\times (1+r_{2n}+r_{3n})
,\]
(here the extinction variation at both
sides of the 'narrow' break is negligible), $r_{2n}=\frac{F_{\lambda ,cont.,{\rm young}}(3933\ \AA )}{F_{\lambda ,cont.,{\rm old}}(3933\ \AA )}$, $r_{3n}=\frac{F_{\lambda ,cont.,{\rm AGN}}(3933\ \AA )}{F_{\lambda ,cont.,{\rm old}}(3933\ \AA )}$.
Using $D_{n,\rm young }(4000)[{\rm 10\ Myr}]=0.96$, 
$D_{\rm n,AGN}(4000)=0.91$ and 
the average values of the parameters
$r_{2n}=0.09^{+0.20}_{-0.05}$, $r_{3n}=0^{+0.38}{-0}$ 
as obtained from the best SED fit of the average galaxy
[Table \ref{Tab:bestfits} [SAB], Fig. \ref{Fig:sedrstacked} upper panel;
$r_{2n}\approx 3\times A_2\times 10^{-0.4\times 
0.40A_V}$, $r_{3n}\approx 3\times A_3\times 10^{-0.4\times 0.40(A_{V,AGN}-A_V}$)],
we get $D_{n,\rm old}(4000)=1.09^{+0.11}_{-0.09}$;
this leads, through the relationship of Fig. \ref{Fig:D4000}/bottom panel, to
$age_{\rm old}=280^{+210}_{-240}$ Myr (68\% CL),$^{+500}_{-280}$ (95\% CL)
 for the stacking of the 13 galaxies;
and $age_{\rm old}=520^{+540}_{-260}$ Myr (68\% CL),$^{+1920}_{-450}$ (95\% CL)
 for the stacking of the four galaxies with F277W-F444W$>$1.7.
Table \ref{Tab:age_sp} summarizes the ranges of ages derived with $D_n(4000)$, which provides better constraints
than other indicators. 

\end{description}

\begin{figure}
\vspace{0cm}
\centering
\includegraphics[width=8cm]{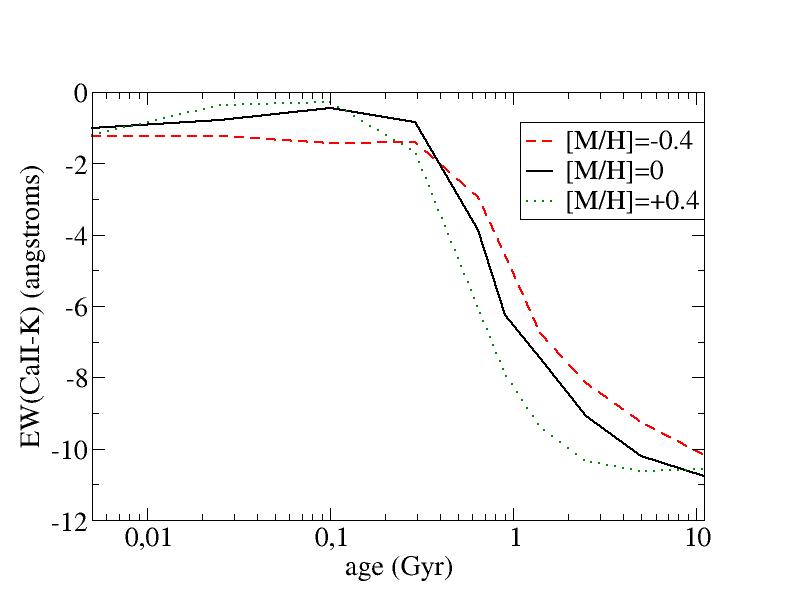}
\vspace{.2cm}
\caption{Prediction of the equivalent width of CaII-K-3933 \AA \ in absorption (negative flux) line for a single
stellar population in GALAXEV stellar population synthesis model, as a function of age and metallicity.}
\label{Fig:EWCaIIK}
\end{figure}



\subsection{CaII index and EW(CaII-K)} 

The ratio [CaII-H(3970)+H$_\epsilon $(in absorption)]/CaII-K(3933) might be used for 
age determination \citep{Leo96,Wil07} in stellar populations. 
However, given the contamination of ongoing star formation or
nuclear activity that causes emission line filling of H$_\epsilon $, the index cannot be directly used as age-dating method \citep{Leo96}. 

The CaII index as 
$\frac{F_\lambda ([3918-3948])-[H_\epsilon ])}{F_\lambda ([3955-3985])}$ can be calculated, 
with $F_\lambda $ corresponding to the flux
once the continuum is subtracted, and the continuum is evaluated at wavelengths 3903-3918 \AA \ and 3985-4000 \AA \ respectively for K and H CaII lines; 
emission flux can be obtained as H$_\epsilon = 0.34H_\gamma 10^{-0.4\times 0.235\,A_V}$, where the intrinsic ratio
0.34 with respect to the emission in H$_\gamma $ stems from \citep[Fig. 1]{Boh15} and the
second factor is the extinction correction. 
As shown in the numbers of the 6th-column of Table \ref{Tab:spectra}, this has very large errors even for the stacked
spectrum of the 13 galaxies: $2.9\pm 2.1$; 
which makes it not useful for any estimation of the age.

None the less, the CaII-K(3933) is also age dependent and can be better constrained.
The value of the equivalent with of CaII-K(3933) for the stacked spectrum of the 13 galaxies is 
-4.7$\pm $2.2 \AA (negative sign means absorption). 
This can be compared with the model predictions as a function of the age, 
once the young component is subtracted (assuming again two SSPs+1 AGN[this does not produce CaII absorption lines]):
\begin{equation}
EW(CaII-K)_{\rm old}=
\end{equation}\[
-r_{2n}\,EW(CaII-K)_{\rm young}
\]\[
+EW(CaII-K)_{\rm old}(1+r_{2n}+r_{3n})
,\]
where $EW(CaII-K)_{\rm old}\equiv EW_{1SSP}(CaII-K)[age_{\rm old},[M/H]]$, 
$EW(CaII-K)_{\rm young}\equiv EW_{1SSP}(CaII-K)[age_{\rm young},[M/H]]=-0.90$ for 10 Myr and
negligible dependence on metallicity. With the previous values of $r_{2n}=0.09^{+0.20}_{-0.05}$, 
$r_{3n}=0^{+0.38}{-0}$, 
$EW(CaII-K)_{\rm old}=-5.1^{+2.4}_{3.1}$ \AA (68\% CL), that means (see Fig. \ref{Fig:EWCaIIK}) 
$age_{\rm old}=800^{+1800}_{-450}$ Myr (68\% CL); no constraint within 95\% CL.
For the four galaxies with F277W-F444W$>$1.7, no constraint is obtained given the
negligible signal-to-noise of the line.

Regrettably, the measurement of the equivalent width of CaII-K line 
for the stacking spectrum is still not accurate enough to provide a tight constraint of the age, but
it is compatible with previous estimates from SED fits.
It is not accurate enough with the present data, but this method might be used in the future with better spectra.

\subsection{Spectral classification and emission lines associated with star formation}
\label{.emlines}

In the average stacked spectrum, the H$_\beta $ line shows an average full width at half-maximum (FWHM) of 1800 km/s after correction for instrumental broadening (calculated by quadratically subtracting the contribution measured from the [OIII] forbidden lines; the small contribution of gas dispersion/rotation speed in the host galaxy is considered negligible in comparison with this number of 1800 km/s in a quadratic sum). This value is consistent with the presence of a moderate AGN component of the Seyfert 1 type. Typically, lines with FWHM$>$1000 km/s are classified as Seyfert 1, while those with FWHM$<$2000 km/s fall into the narrow-line Seyfert 1 (NLS1) category \citep{Goo89}. Among the sample, only two galaxies --\#C-38094 and \#J-1004685-- exhibit significantly broader H$_\beta $ lines, with FWHM$>$2500 km/s. Notably, the first of these is also identified by W24 as a massive Seyfert 1 AGN. 
Overall, the (average) galaxy can be considered NLS1 (or moderate broadening
Seyfert 1). The physical reason for obtaining FWHM$>$1000 km/s is having some (average) moderate mass black hole, 
or better to say that only few of the 31 galaxies have an important Seyfert 1 component with a massive black hole whereas most of the galaxies have none.

For the stacked spectrum of all of the 13 galaxies, 
$\log _{10}\left[\frac{F_\lambda (OIII-5007)}{F_\lambda(H_\beta)}\right]=0.29$, $\log _{10}\left[\frac{F_\lambda (NeIII-3869)}{F_\lambda(H_\beta)}\right]=-0.80$ (without extinction correction). Assuming narrow emission lines, this corresponds to a dominant
component of HII galaxy \citep{Rol97}. Other spectral features also confirm this classification according to 
\citet{Rol97}: $\log _{10}\left[\frac{F_\lambda (OII-3727)}{F_\lambda(H_\beta)}\right]=-1.33$,
$\log _{10}\left[\frac{F_\lambda (OII-3727)}{F_\lambda (NeIII-3869)}\right]=-0.52$, 
$\log _{10}[EW(OII-3727)]=0.98$, $\log _{10}[EW(H_\beta )]=2.14$, $\frac{EW(OII-3727)}{EW(H_\beta )}=0.07$;
$\frac{F_\lambda (H_\beta )}{F_{\lambda,2300\ {\rm \AA}}}=1.07\times 10^2\times 10^{-0.4\times 0.893\,A_V}$ \AA \ (including extinction correction).
These HII regions are associated to the young population necessary to fit the UV-at rest wavelengths.
Similar conclusions are obtained in other analyses of {\it JWST} high-{\it z} galaxies with extreme emission lines \citep[e.g.,][]{Lle24}.

The ratio of emission lines with respect to far-UV-at rest flux, 
associated to  HII regions, cannot be determined
accurately for each galaxy when there are not good SNR spectra, 
but the average values derived from the stacked spectrum of 13 galaxies can be assumed, corrected
of extinction assuming $A_V=0.8$ (as derived in \S \ref{.hbeta}), and assuming $H_\alpha /H_\beta =2.85$: 
The main lines have $F_{\lambda, H_\beta }=(55\ {\rm \AA })\times F_{\lambda,2300\ {\rm \AA}}$; 
$F_{\lambda, Ly\alpha }=1.10F_{\lambda, H_\beta }$,
$F_{\lambda, OII}=0.07F_{\lambda, H_\beta }$, $F_{\lambda, OIIIa}=0.74F_{\lambda, H_\beta }$,
$F_{\lambda, OIIIb}=1.89F_{\lambda, H_\beta }$, $F_{\lambda, H_\alpha }=2.85F_{\lambda, H_\beta }$.
These emission lines will be applied for the models in the SED fits of Sect. \ref{.phot}.

These ratios of emission lines and far-UV-at rest flux are higher (except for OII-line) 
than the usual values observed in low-$z$ galaxies.\footnote{For instance, \citet{Ilb06} gives 
$F_{\lambda, Ly\alpha }=F_{\lambda, OII}$, $F_{\lambda, H_\beta }=0.61F_{\lambda, OII}$, $F_{\lambda, OIIIa}=0.13F_{\lambda, OII}$, $F_{\lambda, OIIIb}=0.36F_{\lambda, OII}$, $F_{\lambda, H_\alpha }=1.77F_{\lambda, OII}$; and
 $F_{\lambda, OII}=(17.6\ {\rm \AA })\times F_{\lambda,2300\ {\rm \AA}}$ 
[derived from the ratios of $F_\nu $ given by \citet{Ken98}, assuming $\lambda =2300$ \AA \ representative
of the average flux per unit wavelength in the range 1500-2800 \AA \ \citep{Ilb06}].} 
The higher ratios observed in the spectra are possibly associated with the peculiarities
of this star formation at high-$z$, as observed by \citet{Mey24}, who suggest burstier star-formation histories and/or more heterogeneous metallicity and ionizing conditions in  $z>7$  galaxies.

\begin{table*}
\caption{Fluxes $F_\lambda$ of some lines (defined positive in emission and negative in absorption),
extinction $A_V$ from the ratio of H$_\beta $ and H$_\gamma $, equivalent width of the line
Ca-II (3933 \AA ), CaII index, and breaks 
$D(4000)$, $D_n(4000)$.  
'all' indicates weighted average stacking of the 13 galaxies;
'colourxxxx' indicates weighted average stacking of the galaxies with the colour 
$C\equiv $F277W-F444W [AB magnitudes] condition xxxx; ($n$g) indicates a number $n$ of galaxies in the sub-sample.
Units of columns 2, 3 and 5: \AA $_{\rm rest}$.}
\begin{center}
\begin{tabular}{cccccccccc}
Stacked sp. &  $\frac{F_\lambda (H_\beta)}{F_\lambda(4000\AA)}$ &
 $\frac{F_\lambda(H_\gamma)}{F_\lambda(4000\AA)}$&
 $A_V$ & EW(CaII-K) & CaII-index & $D(4000)$ & $D_n(4000)$ & 
 $\frac{F_\lambda (OIII-5007)}{F_\lambda(H_\beta)}$ &
  $\frac{F_\lambda (NeIII-3869)}{F_\lambda(H_\beta)}$
\\ \hline
all (13 g) & $164\pm 14$ & $69\pm 7$ & $0.8\pm 1.7$ & $-4.7\pm 2.2$ 
& $2.9\pm 2.1$ & $1.43\pm 0.13$  & $1.08\pm 0.08$  & $1.93\pm 0.24$ & $0.16\pm 0.03$ \\ 
$1.00\le C<1.15$(3g) & $157\pm 37$ & $53\pm 20$ & $2.7\pm 6.8$ & $-7.8\pm 10.3$ 
& $-1.2\pm 2.4$ & $0.89\pm 0.20$  & $1.16\pm 0.36$  & $4.0\pm 1.0$ & $0.36\pm 0.11$ \\ 
$1.15\le C<1.35$(3g) & $72\pm 23$ & $81\pm 28$ & $-7.5\pm 5.8$ & $-29.8\pm 33.6$
& $5.4\pm 4.5$ & $1.12\pm 0.32$ & $4.57\pm 8.06$    & $5.8\pm 1.5$ & $0.35\pm 0.26$ \\ 
$1.35\le C<1.70$(3g) & $199\pm 30$ & $42\pm 24$ & $6.5\pm 10$ & $-8.0\pm 7.8$
& $-0.4\pm 1.5$ & $1.13\pm 0.15$ & $0.77\pm 0.25$   & $2.36\pm 0.50$ & $0.07\pm 0.04$ \\ 
$1.70\le C<2.40$(4g) & $203\pm 22$ & $83\pm 12$ & $1.5\pm 0.2$ & $1.0\pm 2.4$
& $0.3\pm 1.9$ & $1.74\pm 0.17$ & $1.14\pm 0.10$   & $1.48\pm 0.24$ & $0.17\pm 0.03$ \\ 
\hline
\end{tabular}
\end{center}
\label{Tab:spectra}
\end{table*}

\begin{table}
\caption{Average ages (in Myr) derived from spectroscopic Dn4000 
as a function of the range of F277W-F444W (AB magnitudes). $N$ is the number
of galaxies within this constraint. Error bars of ages within 68\%, 95\% CL respectively.}
\begin{center}
\begin{tabular}{cccc}
F277W-F444W & $N$ & $\langle z\rangle $ 
& $\langle {\rm age}_{\rm old}\rangle _{Dn4000}$ 
\\ \hline 
1.00-2.40 & 13& 7.3$\pm 0.2$ 
& 280$^{+210}_{-240}$$^{+500}_{-280}$ 
\\ 
\hline  
1.70-2.40 & 4 & 6.9$\pm 0.5$ 
& 520$^{+540}_{-260}$$^{+1920}_{-450}$ 
\\ \hline 
\end{tabular}
\end{center}
\label{Tab:age_sp}
\end{table}

\section{Photometric SED fits}
\label{.phot}

\subsection{SED fits with two single stellar populations and one AGN}
\label{.results}

\begin{table*}
\caption{Best-fitting results with two SSPs. Errors represent the limits 
within 95\% CL ($\equiv 2\sigma $) 
(within the resolution of the templates), with maximum likelihood method
(Paper I, Appendix A) for $age_{\rm old}$ and
Avni method (Paper I, Sect. 3.1) for the rest of the parameters.
F277W-F444W is the color in AB-magnitudes.
The ages are expressed in Gyr. Redshifts without
error bars are spectroscopic redshifts, obtained from: \citep{Koc23}: [Koc23]; 
\citep{Fuj23}: [Fuj23]; W24;
https://dawn-cph.github.io/dja/spectroscopy/nirspec/ : [S];
https://archive.stsci.edu/hlsp/jades : [J].
Stacked SED: [SA]: stacked of 31 galaxies (4th column of Table \ref{Tab:stacked}), 16 bins;
Stacked SED: [SAB]: S13 except the points at $\lambda =1000$ \AA \, $\lambda =7499$ \AA \, 14 bins;
[S13]: stacked of 13 galaxies with spectroscopic redshift (5th column of Table \ref{Tab:stacked}), 16 bins;
[S13B]: 14 bins.}
\begin{center}
\begin{tabular}{cccccccccc}
Galaxy ID & F277W-F444W & z & $\log _{10}[{\rm age_{\rm old}}]$ & $\log _{10}[{\rm age_{\rm young}}]$ 
& $A_2$ & $A_V$ & $A_3$ & $A_{V,AGN}$ & $\chi ^2_{\rm red}$ 
\\ \hline
C-2859 & 2.30  & $9.81^{+1.39}_{-4.81}$ & $0.28^{+0.76}_{-0.74}$ & $-2.30^{+3.00}_{-0}$ &  $0.02^{+0.08}_{-0.02}$ &
 $0.03^{+1.97}_{-0.03}$ & $0.00^{+0.40}_{-0}$ & $0.00^{+\infty}_{-0}$ & 4.95 \\
C-7274 & 1.70  & $7.96^{+2.04}_{-1.96}$ & $-0.18^{+1.13}_{-2.12}$ & $-1.60^{+2.30}_{-0.70}$ &  $1.54^{+0.13}_{-1.52}$ &
 $0.64^{+1.36}_{-0.64}$ & $0.20^{+0.80}_{-0.15}$ & $8.03^{+8.00}_{-8.03}$ & 33.93 \\
C-11184 & 1.21  & $6.68^{+0.05}_{-0.08}$ &   $0.27^{+0.77}_{-1.01}$ & $-2.30^{+0.70}_{-0}$ &  $0.45^{+0.32}_{-0.12}$ & 
 $0.93^{+1.07}_{-0.53}$ & $0.00^{+0.50}_{-0}$ & $0.00^{+\infty}_{-0}$ & 1.86 \\
C-13050 & 2.02  & $5.62$ [Koc23] & $0.08^{+0.96}_{-2.38}$ & $-2.30^{+0.48}_{-0}$ &  $1.16^{+0.36}_{-0.75}$ &
 $1.17^{+0.49}_{-0.47}$ & $0.52^{+0.32}_{-0.17}$ & $10.11^{+9.83}_{-3.11}$ & 0.58 \\ 
C-14924 & 1.50  & $8.35$ [W24] &  $0.62^{+0.42}_{-0.74}$ & $-2.30^{+3.00}_{-0}$ &  $0.11^{+0.43}_{-0.11}$ &
 $1.06^{+1.14}_{-1.06}$ & $0.06^{+0.94}_{-0.06}$ & $0.00^{+6.94}_{-0}$ & 1.95  \\ 
C-16624 & 1.13  & $8.56^{+0.56}_{-0.45}$ &   $-0.06^{+1.06}_{-2.24}$ & $-2.30^{+0.70}_{-0}$ &  $0.06^{+0.42}_{-0.06}$ & 
 $3.46^{+0.64}_{-2.46}$ & $0.24^{+0.08}_{-0.04}$ & $0.00^{+3.46}_{-0}$ & 0.70 \\     
C-21834 & 1.66 & $8.10^{+1.90}_{-0.25}$ &   $-0.23^{+1.23}_{-2.08}$ & $-2.30^{+0.70}_{-0}$ &  $0.00^{+0.60}_{-0}$ &
 $1.01^{+0.99}_{-0.61}$ & $0.68^{+0.62}_{-0.63}$ & $10.40^{+11.39}_{-10.40}$ & 0.79 \\ 
C-25666 & 1.70 & $5.70^{+2.30}_{-0.15}$ &   $0.13^{+0.91}_{-2.43}$ & $-2.30^{+0.70}_{-0}$ &  $1.23^{+0.19}_{-0.82}$ &
 $2.35^{+0.65}_{-1.85}$ & $0.55^{+0.09}_{-0.35}$ & $0.00^{+5.65}_{-0}$ & 1.87 \\
C-28984 & 1.62 & $7.09$ [S] &   $0.15^{+0.89}_{-1.20}$ & $-1.60^{+0.60}_{-0.70}$ &  $0.32^{+0.28}_{-0.23}$ &
 $0.04^{+1.96}_{-0.04}$ & $0.08^{+0.92}_{-0.03}$ & $20.00^{+19.96}_{-20.00}$ & 1.10  \\
C-35300 & 1.74 & $7.77$ [Fuj23]  &   $-0.12^{+1.13}_{-2.18}$ & $-2.30^{+3.00}_{-0}$ &  $0.40^{+0.32}_{-0.40}$ &
 $1.33^{+0.77}_{-1.33}$ & $0.33^{+0.67}_{-0.33}$ & $7.48^{+8.07}_{-7.48}$ & 6.35 \\
C-37888 & 1.34 & $5.70^{+2.30}_{-0.13}$ &   $-0.10^{+1.12}_{-2.20}$ & $-2.30^{+0.70}_{-0}$ &  $0.96^{+0.02}_{-0.95}$ &
 $0.60^{+1.40}_{-0.60}$ & $1.04^{+0.02}_{-0.99}$ & $5.13^{+7.80}_{-5.13}$ & 2.02  \\
C-38094 & 1.81 & $6.98$ [W24] &  $-0.18^{+1.02}_{-0.51}$ & $-2.30^{+0.70}_{-0}$ &  $0.34^{+0.44}_{-0.15}$ &
 $1.22^{+0.78}_{-0.28}$ & $0.35^{+0.55}_{-0.30}$ & $6.72^{+7.42}_{-6.72}$ & 6.78  \\
C-39575 & 1.32 & $7.99$ [Fuj23] &  $-0.18^{+1.15}_{-2.12}$ & $-2.30^{+3.00}_{-0}$ &  $0.19^{+0.41}_{-0.19}$ &
 $0.32^{+1.68}_{-0.32}$ & $0.00^{+1.00}_{-0}$ & $2.36^{+\infty}_{-2.36}$ & 2.85  \\ \hline
J-0015355 & 1.13 & $8.13^{+1.87}_{-0.25}$ & $-0.06^{+1.11}_{-2.24}$ & $-2.30^{+3.00}_{-0}$ &  $3.48^{+0}_{-3.45}$ & 
 $0.87^{+1.13}_{-0.87}$ & $0.00^{+1.00}_{-0}$ & $0.00^{+\infty}_{-0}$ & 1.56 \\
J-0066263 & 1.42  & $8.20^{+1.80}_{-0.42}$ & $0.30^{+0.75}_{-2.60}$ & $-2.30^{+1.30}_{-0}$ &  $0.25^{+0.86}_{-0.22}$ & 
 $1.04^{+0.96}_{-1.04}$ & $0.02^{+0.98}_{-0.02}$ & $0.00^{+5.96}_{-0}$ & 2.96 \\
J-0075634 & 1.18 & $8.25^{+1.75}_{-0.44}$ & $0.35^{+0.69}_{-2.66}$ & $-2.30^{+3.00}_{-0}$ &  $1.01^{+0.31}_{-0.98}$ & 
 $0.94^{+1.06}_{-0.94}$ & $0.02^{+0.98}_{-0.02}$ & $5.28^{+7.06}_{-5.28}$ & 4.81 \\
J-0165305 & 1.03 & $7.36^{+0.12}_{-0.19}$ & $-0.07^{+1.11}_{-2.23}$ & $-2.30^{+0.48}_{-0}$ &  $3.48^{+0}_{-3.24}$ & 
 $0.60^{+0.40}_{-0.33}$ & $0.00^{+0.55}_{-0}$ & $0.00^{+\infty}_{-0}$ & 11.48 \\ 
J-0204022 & 1.13 & $6.99^{+0.44}_{-0.15}$ & $-0.72^{+1.67}_{-1.59}$ & $-2.30^{+0.48}_{-0}$ &  $3.48^{+0}_{-3.45}$ & 
 $0.91^{+0.69}_{-0.41}$ & $0.02^{+0.38}_{-0.02}$ & $7.66^{+6.85}_{-7.66}$ & 6.27 \\
J-0210600 & 1.26 & 6.31 [J] & $0.11^{+0.93}_{-2.41}$ & $-2.30^{+0.70}_{-0}$ &  $0.51^{+0.31}_{-0.50}$ &
 $0.95^{+1.05}_{-0.55}$ & $0.11^{+0.89}_{-0.11}$ & $3.77^{+5.45}_{-3.77}$ & 8.75 \\ 
J-0211388 & 1.01 & 8.38 [J] & $0.63^{+0.41}_{-2.93}$ & $-2.30^{+0.70}_{-0}$ &  $1.60^{+0.15}_{-1.36}$ &
 $0.60^{+1.40}_{-0.60}$ & $0.30^{+0.20}_{-0.30}$ & $4.93^{+7.80}_{-4.93}$ & 2.06 \\ 
J-0214552 & 1.23 & $5.67^{+0.36}_{-0.08}$ & $-1.60^{+2.64}_{-0.70}$ & $-2.30^{+0.70}_{-0}$ &  $0.00^{+0.60}_{-0.00}$ &
 $1.62^{+0.48}_{-1.12}$ & $0.22^{+0.78}_{-0.12}$ & $0.00^{+2.38}_{-0}$ & 2.00 \\ \hline 
J-1001830 & 1.00 & 6.67 [J] & $-0.31^{+1.29}_{-1.99}$ & $-2.30^{+0.48}_{-0}$ &  $3.48^{+0}_{-3.45}$ &
 $1.11^{+0.89}_{-0.47}$ & $0.44^{+0.56}_{-0.39}$ & $20.00^{+18.89}_{-20.00}$ & 0.50 \\
J-1004685 & 2.27  & 7.42 [J] & $-0.35^{+1.38}_{-0.73}$ & $-2.30^{+3.00}_{-0}$ &  $0.18^{+0.42}_{-0.18}$ &
 $0.52^{+1.48}_{-0.52}$ & $1.28^{+0.04}_{-1.23}$ & $9.87^{+11.88}_{-9.87}$ & 0.91 \\
J-1010260 &  1.46 & 8.28 [J] & $0.37^{+0.68}_{-2.67}$ & $-2.30^{+3.00}_{-0}$ &  $0.36^{+0.24}_{-0.33}$ &
 $0.00^{+2.00}_{-0}$ & $2.20^{+0}_{-2.15}$ & $13.71^{+14.64}_{-13.71}$ & 2.15 \\ 
J-1010816 &  1.31 & 6.76 [J] & $0.40^{+0.59}_{-0.64}$ & $-2.30^{+0.70}_{-0}$ &  $0.33^{+0.29}_{-0.22}$ &
 $0.72^{+1.28}_{-0.72}$ & $0.00^{+1.00}_{-0}$ & $0.00^{+\infty}_{-0}$ & 1.88 \\
J-1024087 &  1.26 & $8.03^{+1.97}_{-3.03}$ & $-0.37^{+1.30}_{-1.93}$ & $-2.30^{+3.00}_{-0}$ &  $0.27^{+0.33}_{-0.27}$ &
 $0.00^{+2.00}_{-0}$ & $0.00^{+1.00}_{-0}$ & $0.00^{+\infty}_{-0}$ & 2.51 \\ 
J-1032447 &  1.14 & 7.09 [J] & $0.12^{+0.92}_{-2.42}$ & $-2.30^{+0.70}_{-0}$ &  $3.48^{+0}_{-3.36}$ &
 $0.76^{+1.14}_{-0.76}$ & $0.30^{+0.70}_{-0.30}$ & $8.46^{+10.64}_{-8.46}$ & 2.73 \\
J-1042550 &  1.56 & $7.21^{+0.32}_{-0.13}$ & $0.11^{+0.93}_{-2.41}$ & $-2.30^{+0.48}_{-0}$ &  $3.48^{+0}_{-2.85}$ & 
$1.14^{+0.86}_{-0.44}$ & $0.00^{+0.26}_{-0}$ & $0.00^{+\infty}_{-0}$ & 3.78 \\
J-1043804 & 1.11 & $5.71^{+0.36}_{-0.12}$ & $-0.66^{+1.43}_{-1.64}$ & $-2.30^{+0.48}_{-0}$ &  $1.60^{+0.50}_{-1.22}$  & 
$0.68^{+1.32}_{-0.46}$ & $0.03^{+0.32}_{-0}$ & $20.00^{+19.32}_{-20.00}$ & 1.34 \\
J-1050323 & 1.72 & $6.90^{+0.03}_{-0.05}$ & $0.98^{+0.06}_{-0.39}$ & $-2.30^{+0.48}_{-0}$ &  $0.34^{+0.29}_{-0.13}$ & 
 $0.92^{+1.08}_{-0.28}$ & $0.00^{+0.75}_{-0}$ & $0.00^{+\infty}_{-0}$ & 5.82 \\ 
J-1061176 & 1.49 & $8.19^{+1.81}_{-0.31}$ & $0.07^{+0.97}_{-2.37}$ & $-2.30^{+0.70}_{-0}$ &  $1.39^{+0.24}_{-1.36}$ & 
 $1.06^{+0.94}_{-1.06}$ & $1.28^{+0}_{-1.23}$ & $13.80^{+13.58}_{-13.80}$ & 1.77 \\
\hline  \hline
SA  & --- & --- & $-0.32^{+0.34}_{-0.71}$ & $-2.30^{+2.26}_{-0}$ &  $0.03^{+0.54}_{-0.03}$ &
 $0.00^{+2.00}_{-0.00}$ & $0.00^{+0.50}_{-0.00}$ & $0.00^{+\infty }_{-0}$ & 3.89  \\ 
S13  & --- & --- & $-0.76^{+1.31}_{-1.54}$ & $-2.30^{+3.00}_{-0}$ &  $3.48^{+0}_{-3.45}$ &
 $3.15^{+0.01}_{-3.15}$ & $0.52^{+0.48}_{-0.52}$ & $0.00^{+3.85}_{-0}$ & 4.06  \\ 
SAB  & --- & --- & $-0.32^{+0.24}_{-0.70}$ & $-2.30^{+2.11}_{-0.00}$ &  $0.03^{+0.13}_{-0.03}$ &
 $0.00^{+1.50}_{-0.00}$ & $0.00^{+0.25}_{-0.00}$ & $0.00^{+\infty }_{-0}$ & 1.98  \\ 
S13B  & --- & --- & $0.10^{+0.92}_{-1.24}$ & $-0.05^{+0.74}_{-2.26}$ &  $1.06^{+0.02}_{-1.06}$ &
 $0.00^{+2.00}_{-0.00}$ & $0.22^{+0.13}_{-0.14}$ & $0.00^{+4.02}_{-0}$ & 1.39  \\ 
SAB, $A_3=1$ & --- & --- & $-0.34^{+1.19}_{-1.96}$ & $-2.30^{+2.26}_{-0.00}$ &  $3.48^{+0.00}_{-3.45}$ &
 $1.90^{+0.20}_{-1.90}$ & $1.00$ & $2.32^{+1.20}_{-1.22}$ & 8.45 \\ 
\hline
\end{tabular}
\end{center}
\label{Tab:bestfits}
\end{table*}

Following the method of Paper I (see also \citet{Lop17,Gao24}), 
the photometric fluxes of the 31 galaxies are fitted 
to GALAXEV model \citep{Bru03} with two
single stellar populations (SSPs) plus one AGN component, with a total of eight or nine free pararameters: 
age and luminosity amplitudes of the young and old populations ($A_2$ is the ratio of young and old population luminosity
at 5500 \AA -rest; there is not any constraint in the range of ages, we just require that $age_{\rm young}<age_{\rm old}$), 
average interstellar extinction $A_V$, metallicity, 
$A_3$ representing the ratio of AGN/old rest luminosity at 5500 \AA \ , $A_{V,AGN}$ as the extinction of the AGN component,
and redshift (only when spectroscopic redshift is not available).

Dust emission is neglected at $\lambda _{\rm rest}\lesssim$ 2$\mu $m for stellar components, but taken
into account in the AGN component.
Emission lines of HII regions associated with the young populations are 
included by assuming the ratios derived in \S \ref{.emlines}: 
$F_{\lambda, H_\beta }=(55\ {\rm \AA })\times F_{\lambda,2300\ {\rm \AA}}$; $F_{\lambda, Ly\alpha }=1.10F_{\lambda, H_\beta }$, $F_{\lambda, OII}=0.07F_{\lambda, H_\beta }$, $F_{\lambda, OIIIa}=0.74F_{\lambda, H_\beta }$,
$F_{\lambda, OIIIb}=1.89F_{\lambda, H_\beta }$, $F_{\lambda, H_\alpha }=2.85F_{\lambda, H_\beta }$.
\footnote{An error in the code 
used in Paper I was found, by which emission lines were not included in their fits in
spite of their claim that emission lines were included.  
The effect of including emission lines of HII regions associated with young population
with usual values of emission lines and far-UV-at rest flux as given by \citet{Ken98} and \citet{Ilb06}
would be negligible, much smaller than the error bars. A higher ratio of emission lines, as used here,
gives more significant deviations, though still compatible with previous ones within their error bars.\\
Note that these ratios of emission lines reflect average properties of HII regions, and there may be 
a significant scatter of their values, which is not taken here into account.
These and other (e.g., NeIII) strong lines typical of AGN are included in the template of AGN.}

The best-fitting results for the 31 galaxies in the sample are shown in 
Table~\ref{Tab:bestfits}. Most cases are compatible 
(within 95\% CL) with zero extinction. The metallicity ([M/H] between -0.4 and +0.4; 
not included in the Table)
has very small effect in the fits, and any value
of it fits the SED distribution within the error bars.

The average redshift of these 31 galaxies 
is $\langle z\rangle =7.35\pm 0.19$ (1$\sigma $; corresponding to rms=1.04), at which the $\Lambda$CDM universe 
had an age $t_{\rm Univ}(\langle z\rangle)=700^{+30}_{-20}$ (1$\sigma $) Myr.
For the 13 galaxies with spectra, the average redshift is 7.29, very similar to the photometric+spectroscopic 
sample.

The weighted average (see Appendix B of Paper I; with the error bars including the precision
of the age in the used templates of GALAXEV)
of $\log _{10}({\rm age}_{\rm old})$ of these galaxies, converted to linear scale, gives 
\begin{equation}
\label{av1}
\langle {\rm age}_{\rm old}\rangle =1.2
^{+0.5}_{-0.4}\;(95\% {\rm CL})\ {\rm Gyr} 
.\end{equation}
For the sample of 31 galaxies, within the error bars, the average age of the $\Lambda $CDM universe 
can be larger than the average age of these galaxies of Eq. (\ref{av1}).

\begin{figure}[h]
\vspace{0cm}
\centering
\includegraphics[width=8cm]{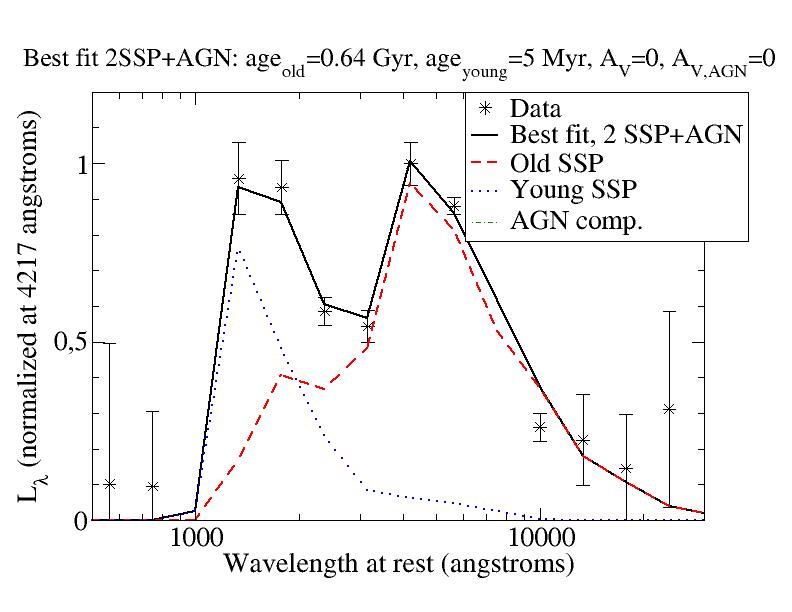}
\vspace{.2cm}
\includegraphics[width=8cm]{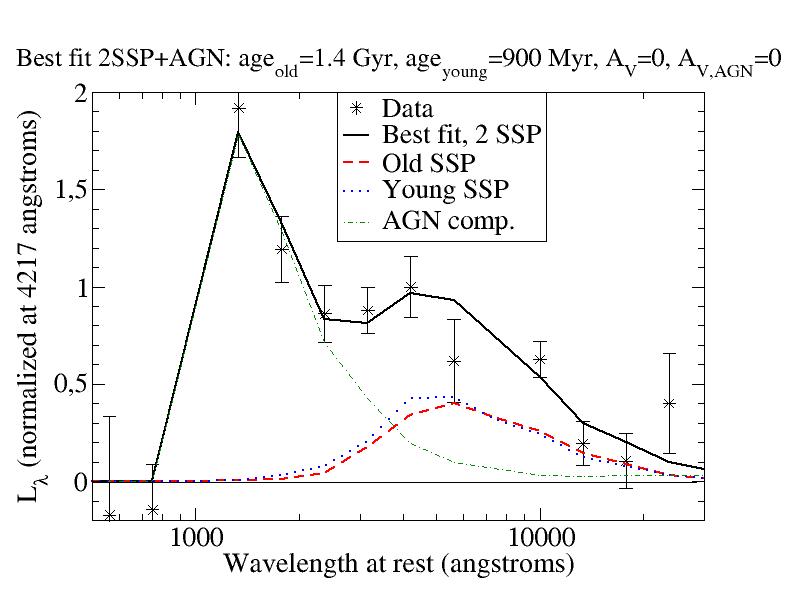}
\vspace{.2cm}
\caption{Top: Linear-log plot of the best fit using two SSPs + one AGN of the stacked SED at rest of all galaxies [SAB] ($\chi ^2_{\rm red}=1.98$); or
bottom: only the 13 galaxies with spectroscopic redshift [S13B] ($\chi ^2_{\rm red}=1.39$), excluding 
the bins at 1000 and 7499 \AA . Data in Table \ref{Tab:stacked}.}
\label{Fig:sedrstacked}
\end{figure}

\begin{table}
\caption{Stacked SED at rest for the 31 galaxies, and for the 13 galaxies 
with spectroscopic redshift. Fluxes are normalized to unity at 4217 \AA . 
$N$ or $N_{\rm sp}$  is the number of points that have some contribution 
in the bin with all or spectroscopic galaxies.}
\begin{center}
\begin{tabular}{ccccc}
$\lambda _{\rm rest}$(\AA ) & $N$ & $N_{\rm sp}$ & $\langle F_\lambda \rangle$ (all) & 
$\langle F_\lambda \rangle$ (spec. gal.) \\ \hline
      422 & 14 & 8       & 1.097$\pm $2.429 & 0.308$\pm $2.443 \\  
      562 & 24 & 16      & 0.104$\pm $0.393 & -0.171$\pm $0.508 \\
      750 & 40 & 24      & 0.094$\pm $0.211 & -0.145$\pm $0.234 \\
     1000 & 59 & 29      & 0.520$\pm $0.198 & 1.179$\pm $0.430 \\
     1334 & 59 & 28      & 0.958$\pm $0.101 & 1.918$\pm $0.252 \\
     1778 & 61 & 26      & 0.933$\pm $0.076 & 1.193$\pm $0.171 \\
     2371 & 49 & 21      & 0.586$\pm $0.038 & 0.862$\pm $0.148 \\
     3162 & 44 & 22      & 0.545$\pm $0.046 & 0.879$\pm $0.118  \\
     4217 & 58 & 28      & 1.000$\pm $0.060 & 1.000$\pm $0.157  \\
     5623 & 34 & 17      & 0.881$\pm $0.023 & 0.619$\pm $0.214  \\
     7499 & 17 & 10      & 0.252$\pm $0.083 & 1.952$\pm $0.181  \\
    10000 & 17 & 9       & 0.262$\pm $0.039 & 0.627$\pm $0.094  \\ 
    13335 & 7  & 4       & 0.223$\pm $0.128 & 0.195$\pm $0.112  \\
    17783 & 8  & 3       & 0.147$\pm $0.153 & 0.105$\pm $0.142  \\
    23714 & 6  & 2       & 0.312$\pm $0.276 & 0.402$\pm $0.259  \\
    31623 & 1  & 1       & 1.158$\pm $1.784 & 1.341$\pm $1.502  \\
     \hline
\end{tabular}
\end{center}
\label{Tab:stacked}
\end{table}

\subsection{Stacked SED}
\label{.stacked}

A stacked SED is obtained 
summing the 31 SEDs at rest (assuming the redshift inferred from their best fits): [SA]; or
only the 13 SEDs at rest of only the galaxies [S13] with available 
spectroscopic redshift.
The procedure of sect. 3.2 of
Paper I is followed, we are stacking the photometry (at rest and normalized) of the individual galaxies, but doing weighted average 
for each bin, thus the dilution of
high S/N fluxes when mixed with low S/N ones is avoided; with a fixed interval in logarithmic scale of wavelengths
of $\Delta (\log_{10} \lambda)=1/8$. 

The stacked SEDs 
are given in Fig.~\ref{Fig:sedrstacked} 
and Table \ref{Tab:stacked}, 
whose optimized parameters (excluding the redshift, which is fixed at zero in the rest SED;
here there are in total eight free parameters) are 
given in Table~\ref{Tab:bestfits}. 
The IGM extinction is calculated assuming a Gaussian distribution of redshifts $5<z<10$ with average 7.3 and rms=1.0.
The results for both stacked SEDs present V-shape, although
the spectroscopic galaxies show a more conspicuous flux peak around Ly$\alpha $ and H$_\alpha $ lines,
possibly because the spectroscopic galaxies contain higher ratio of star formation or 
AGNs. If the bins at 1000 and 7499 \AA \ are removed, which due to the strong gradients associated to those lines
give poorer constraints, [SAB] and [S13B] respectively, better fits are obtained.

The most interesting parameter of the fit of stacked SED is 
the age of the oldest population corresponding to this average stacked SED [error derived using
maximum likelihood (Paper I, appendix A)], which has tighter constraint for [SAB]:
\begin{equation} 
\label{av3}
=0.48^{+0.13}_{-0.24}(68\% {\rm CL})^{+0.35}_{-0.38} (95\% {\rm CL}) \ {\rm Gyr} \  
.\end{equation}

\begin{figure}
\vspace{0cm}
\centering
\includegraphics[width=8cm]{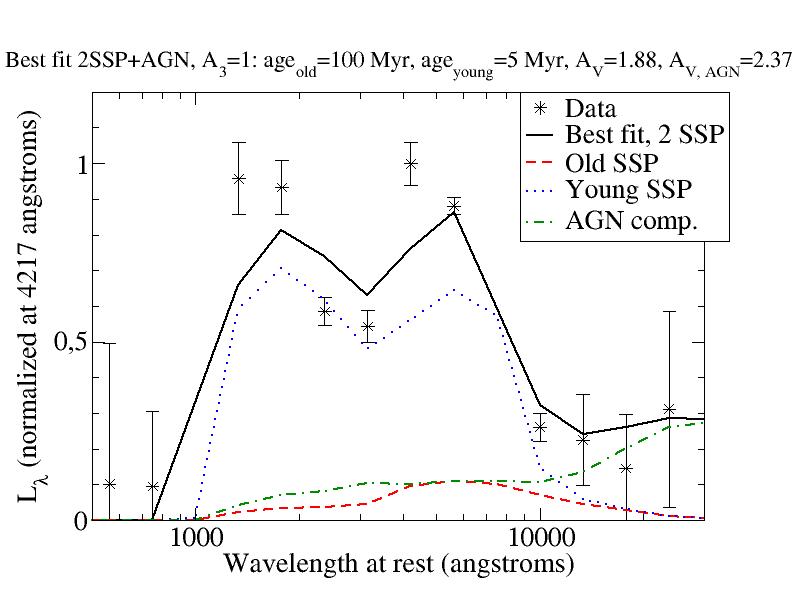}
\vspace{.2cm}
\caption{Best fit of the stacked SED at rest [SAB] including an AGN 
component with $A_3=1$ 
($\chi ^2_{\rm red}=8.39$).}
\label{Fig:sedrstacked4ee}
\end{figure}

In [SAB], the best fit (see Table \ref{Tab:bestfits}) corresponds to $A_3=0$, 
plotted in Fig. \ref{Fig:sedrstacked}/top panel, but within 95\% CL values of $A_3<0.25$ 
are allowed. 
As shown in Fig. \ref{Fig:sedrstacked4ee}, the best solution with $A_3=1$ (equal contribution of AGN and old population
at 5500 \AA ) gives a much worse fit: $\chi ^2_{\rm red}$ more than 4 times larger than with $A_3=0$, because
it requires a dominant young population, which cannot simultaneously fit the Ly-$\alpha $ and Balmer break. 
This means that among the 31 galaxies there may be very few galaxies with
a strong component of AGN, while most of them lack any AGN.

When $A_3>0$ the range of allowed ages for the old population is very similar to that
 with $A_3=0$.
The effect of including an AGN on the age of the stellar populations is very low, at most it reduces 40 Myr
the age of the minimum age of the galaxy within 95\% CL.



The difference between Eqs. (\ref{av1}) and (\ref{av3}) may reflect that the average age of SEDs 
is not the age of average SED, due mainly to some outliers (see \S \ref{.old}) that get higher weight in Eq. (\ref{av1}) than in Eq. (\ref{av3}) (with weight proportional to the flux). They are two different numbers and they do not necessarily overlap, although both reflect some type of average age of the old population with different weights for the different individual galaxies. For an estimation, we consider that an intermediate value better reflects the average age and its dispersion.
The average of both expressions (they do not sum quadratically because they are not independent) is:
\begin{equation}
\langle age_{\rm old}\rangle=0.84\pm 0.41 (95\% {\rm CL})\ {\rm Gyr}
.\end{equation}

\subsection{Far infrared limits}

Assuming the same \citet{Ass10}'s AGN template including dust used here and in Paper I, and neglecting the extinction at far infrared in comparison with $A_{V,AGN}$:
\begin{equation}
\frac{F_{\nu, AGN,25\mu m-rest}}{F_{\nu, V-rest}}\approx 230\times\frac{A_3}{1+A_2+A_3}\times 10^{0.4A_{V,AGN}}
.\end{equation}
V-rest$\equiv $filter F444W of {\it JWST}/NIRCAM (Near-infrared camera) and 25$\mu $m-rest$\equiv $ 250$\mu $m is approximately assumed, 
given an average $z\sim 7.3$, and the average $F_{\nu, F444W}=160$ nJy for the 31 galaxies. 
Hence, for $\frac{A_3}{1+A_2+A_3}<0.2$ and $A_{V,AGN}<4$
$F_{\nu, AGN,250\mu m}\lesssim 0.30$ mJy is got on average for each source.
This is not detectable in any of the available surveys in this frequency \citep[Table 2]{Lop21}.
As a matter of fact, none of the sources is detected in {\it Herschel}/SPIRE-250$\mu $m, but the limit
of detection of a source is 10 mJy \citep{Eal10}. 

At longer wavelengths, $F_{\nu, AGN,140\mu m-rest}\sim 1.5\,F_{\nu, AGN,25\mu m-rest}$ (\citet[Fig. 4]{Cal21};\ \citet{Lab23b}), so one would expect $F_{\nu, AGN,1200\mu m}\lesssim 0.44$ mJy on average with 1.2mm observations.
 This is within the limit of detection in the most powerful submillimeter array interferometers: 
 for instance, the typical error bar in {\it JWST}-ALMA (Atacama Large Millimeter/submillimeter Array; Southern hemisphere) survey sources 
is 40 $\mu $Jy \citep{Lab23b}, so, in case the AGN contribution is $\sim $20\% of the light 
at 5500 \AA \ and the extinction of the AGN is as high as 4 mag, 
some sources with $F_{\nu, AGN,1200\mu m}\gtrsim 0.4$ mJy might be
 detected at 10$\sigma $ level; IRAM (Institut de Radioastronomie Millim\'etrique; Northern hemisphere) may get similar figures.
However, other estimations indicate much lower emissions, impossible to be detected by ALMA \citep{Cas24}, 
indicating that the amount of dust is much lower than here assumed.

On other similar sources at different coordinates, so far, none of this type of JWST extremely red and compact galaxies (LRDs) at $z\gtrsim 6$ in other areas could be detected in ALMA \citep{Lab23b,Wil24}, 
thus rejecting the hypothesis of red QSOs as their dominant component. Also, in a spectroscopic analysis
\citep{Per24}, only 3 out of 18 high-$z$ {\it JWST} candidates of LRDs presented signs of AGN.
Our galaxies were not selected to be LRDs, since we do not include any constraint on angular size, 
but most of them are within this category (see Sect. \ref{.data}).

\subsection{Age dependent on colour}
\label{.color}

The whole sample of 31 galaxies was selected with two color criteria for Ly-$\alpha $ and Balmer breaks:
F150W-F277W$<$0.7, F277W-F444W$>$1.0 in AB magnitudes.
The first colour break ensures a low extinction in the young stellar population, and the second break needs an old age
of the older stellar population. It is expected a dependence of the age on the second color criterion within
the selected constraints.
Therefore, older average of the galaxies can be obtained if the reddest colours are selected.
In Table \ref{Tab:age_vs_color}, the values of the average ages are given, with different estimators 
for different constraints on $C\equiv $F277W-F444W. 
As said previously, the difference between both expressions calculated with the two different methods 
may reflect that the average age of SEDs 
is not the age of average SED.
The dependence of the color is significant
 (within 3$\sigma $) for 
$\langle \log _{10}[{\rm age}_{\rm old,2SSP+AGN}]\rangle $ (4rd column of Table \ref{Tab:age_vs_color}).
This is in agreement with the results by \citet{Mar25}, who finds redder colour for the oldest galaxies.

\begin{table*}
\caption{Average ages (in Gyr) as a function of the range of F277W-F444W (AB magnitudes). $N$ is the number
of galaxies within this constraint. $\langle {\rm age}_{\rm old, 2SSP}\rangle $ is the weighted average
of the fits of individual galaxies with 2SSP. [SAB] indicates
average over the stacked SED excluding the bin containing H$_\alpha$ line; 
fits with 2SSP+AGN. Last column is the average of columns 4, 5 in linear scale. 
Error bars within 95\% CL. }
\begin{center}
\begin{tabular}{cccccc}
F277W-F444W & $N$ & $\langle z\rangle $ & $\langle \log _{10}[{\rm age}_{\rm old,2SSP+AGN}]\rangle $  
 & $\log_{10}[{\rm age}_{\rm old;SAB,2SSP+AGN}]$ & $\langle {\rm age}\rangle $ (Myr)
  \\ \hline 
1.00-2.40 & 31& 7.35$\pm 0.19$ &  0.08$^{+0.15}_{-0.17}$ &  -0.32$^{+0.24}_{-0.70}$  
& $840\pm 410$
\\ \hline 
\hline 
1.00-1.15 & 8 & 7.36$\pm 0.34$ &  -0.40$^{+0.24}_{-0.29}$ &  -0.56$^{+1.16}_{-1.74}$ & $340^{+2000}_{-240}$
\\   
1.15-1.35 & 8 & 6.92$\pm 0.37$ &  -0.02$^{+0.29}_{-0.34}$ &  -0.59$^{+1.04}_{-0.58}$ & $600^{+1740}_{-300}$
 \\  
1.35-1.70 & 7 & 7.92$\pm 0.20$ &  0.13$^{+0.19}_{-0.26}$ &  0.07$^{+0.97}_{-2.37}$ & $1260^{+5260}_{-890}$
 \\ 
1.70-2.40 & 8 & 7.27$\pm 0.47$ &  0.24$^{+0.34}_{-0.37}$ &  -0.07$^{+1.11}_{-1.09}$ & $1300^{+6100}_{-890}$
 \\ 
 \hline
\end{tabular}
\end{center}
\label{Tab:age_vs_color}
\end{table*}

\subsection{Oldest galaxies}
\label{.old} 

\subsubsection{Three galaxies with $age_{\rm old}\ge 600$ Myr (95\% CL)}
 
In the sample of 31 galaxies, our SED fits (Table \ref{Tab:bestfits}) indicate that there are three galaxies
with age$_{\rm old}\ge 600$ Myr (95\% CL): C-14924, J-1010816, and J-1050323. One of these galaxies even gets age$_{\rm old}>3$ Gyr (95\% CL). In Fig. \ref{Fig:SEDgals} we show the respective SED with the corresponding best fits.

\begin{itemize}
\item
C-14924 gets a SED fit constraint of age$_{\rm old}=4.2^{+6.8}_{-3.4}$(95\%CL,max.likelihood) Gyr, that is, no
combination of old population + young population + AGN and corresponding extinction allows
age$_{\rm old}<800$ Myr within 95\% CL. The SED shows a
conspicuous V-shape with a strong increase of flux after Balmer break. 
AGN contribution is small in the red part for the SED, though
allowing any value $A_3<1.0$ within 95\% CL. 
Luminosity:
$L_{\rm V,rest}=2.3\times 10^{10}$ L$_\odot $ ($M_V=-21.1$).
There is a spectrum for this galaxy,
which reveals narrow H$_\beta $ line; $\frac{F_\lambda (OIII-5007)}{F_\lambda(H_\beta)}=1.48\pm 0.11$; 
$\frac{F_\lambda (NeIII-3869)}{F_\lambda(H_\beta)}=0.06\pm 0.04$; typical of a HII-region with negligible AGN
contribution (in the red part of the spectrum).

\item
J-1010816 gets a SED fit constraint of age$_{\rm old}=2.5^{+7.3}_{-1.9}$(95\%CL,max.likelihood) Gyr, that is, no
combination of old population + young population + AGN and corresponding extinction allows
age$_{\rm old}<600$ Myr within 95\% CL. Thanks to the intermediate-width filters in JADES, we see the
imprint of H$_\beta $+OIII lines in the photometric data around 5000 \AA , which is correctly reproduced
by the contribution of the young population. AGN contribution is zero or negligible for the best fit, though
allowing any value $A_3<1.0$ within 95\% CL. 
Luminosity:
$L_{\rm V,rest}=1.4\times 10^{10}$ L$_\odot $ ($M_V=-20.5$).
There is also a spectrum for this galaxy,
which reveals narrow H$_\beta $ and H$_\alpha $ lines
; $\frac{F_\lambda (OIII-5007)}{F_\lambda(H_\beta)}=2.78\pm 0.16$; 
$\frac{F_\lambda (NeIII-3869)}{F_\lambda(H_\beta)}=0.23\pm 0.04$; typical of a HII-region with negligible AGN
contribution (in the red part of the spectrum).

\item
J-1050323 is the most outrageous case.
Remarkably, the relative error bars of the photometric points in this SED are very low: between 3500 and 6000 \AA \ at rest, they are less than 2\%; the mid-infrared points are also quite good. This allows a quite precise constraint.
It gets a SED fit constraint of age$_{\rm old}=9.5^{+1.5}_{-5.6}$(95\%CL,max.likelihood) Gyr, that is, no
combination of old population + young population + AGN and corresponding extinction allows
age$_{\rm old}<3.9$ Gyr within 95\% CL. Also, $<1.8$ Gyr is excluded within 99.7\% CL (3$\sigma$ )
and $<0.8$ Gyr (the age of the Universe at the corresponding redshift $z=6.90$) has a probability
$<3\times 10^{-7}$ ($>5\sigma $).
  Again, with the intermediate-width filters in JADES, we see the
imprint of H$_\beta $+OIII lines in the photometric data around 5000 \AA , which is correctly reproduced
by the contribution of the young population. AGN contribution is zero or negligible for the best fit, though
allowing any value $A_3<0.75$ within 95\% CL. 
High luminosity:
$L_{\rm V,rest}=3.1\times 10^{10}$ L$_\odot $ ($M_V=-21.4$).
There is no spectrum available for this galaxy.

\end{itemize}

\subsubsection{Effect of emission lines}
\label{.oldemlines}

In all of the calculations of the age of these three galaxies, we have assumed a SED fit including emission lines corresponding to
the ratios in the average spectrum (\S \ref{.emlines}), with star formation ratio with respect to UV emission associated to the young
population indicated by $R_{\beta-UV}=
\frac{F_{\lambda, H_\beta }}{F_{\lambda,2300\ {\rm \AA}}}=55$ \AA. For the SED fit of each individual galaxy, 
there is a systematic error due to the uncertainty of
the true value of $R_{\beta-UV}$, which may vary with respect to the average value in the 13 spectroscopic galaxies.
In order to estimate this systematic error of age$_{\rm old}$ produced by this variation of the star formation ratio in the most critical case, 
the galaxy J-1050323, we carry out the calculation with free $R_{\beta-UV}$ (thus, we have 10 free parameters instead of 9). The best fit gives $\chi _{\rm red}^2=0.99$, for $R_{\beta-UV}=90^{+80}_{-20}$ \AA (95\%CL), age$_{\rm old}=10.2^{+0.8}_{-8.1}$(95\%CL,max.likelihood) Gyr. An age of the galaxy $<0.8$ Gyr (the age of the Universe at the corresponding redshift $z=6.90$) has a probability $<1.5\times 10^{-6}$ ($>4.7\sigma $). 
The obtained age range is similar to the previous result with constant $R_{\beta-UV}=55$ \AA . 
Therefore, the problem of the age of J-1050323 much larger than the age of the Universe is not solved by taking account this uncertainty in the ratio of emission lines, which is a minor factor.           

\subsubsection{TP-AGB effect}
\label{.TP-AGB}

TP-AGB (thermally pulsing asymptotic giant branch) phase was not considered or very lightly 
in the GALAXEV model \citep{Bru03} used here.

The model of \citet{Mar05} is pioneer in the calculation of this effect, and the comparison of this model with GALAXEV
\citep[Fig. 19]{Mar05} show important differences. 
TP-AGB continous emission enhancement at $\lambda _{\rm rest} >$ 7000 \AA \ is only relatively important 
for ages larger than 0.6-1.0 Gyr \citep{Mar05,Lop17,Mar25,Lu25}.
More recent versions of the Maraston model \citep{Mar20} do not give important variations.

\citet{Mar05} differences with GALAXEV reflect the TP-AGB effect, but also
possible differences due to different initial mass function (IMF) 
or other parameters in the stellar evolution models. 
Instead, for the comparison here we will use a 2007 revised version of GALAXEV model [BC07, by Charlot \& Bruzual
2007\citep{Bru07}], also including similar enhanced TP-AGB effect, very slighly weaker.
We will estimate the effect of including a TP-AGB by introducing an approximation in the comparison of the models:

\begin{equation}
R_{\rm TP-AGB}(\lambda ,{\rm age})=\frac{L_{\lambda , {\rm CB07}}({\rm age})}{L_{\lambda ,{\rm GALAXEV}}({\rm age})}\approx
1+F_{1.25\mu m}({\rm age})\,H_{1{\rm Gyr}}(\lambda )
\end{equation}\[
H_{1{\rm Gyr}}(\lambda )=\left \{ 
\begin{array}{ll}
0,& \mbox{$\lambda <7000$ \AA} \\
0.96l-0.173l^2,& 
\mbox{$7000$ \AA $\le \lambda \le 40000$ \AA} 
\end{array}
\right \} \;.
\]\[
F_{1.25\mu m}({\rm age})=1.36\times \exp \left[-3.45|\log_{10}({\rm age}({\rm Gyr}))+0.09|\right]
,\]
where $l\equiv \left(\frac{\lambda}{7000\ {\rm A}}-1\right)$.
We have neglected here the dependence on the metallicity, taking the solar one as the reference. 
Spectral features are not contemplated here, but they
are not important for photometry with intermediate-large wide filters. 

Introducing this TP-AGB enhancement (that is, multiplying $L_\lambda $ of each SSP of the GALAXEV model by 
$R_{\rm TP-AGB}(\lambda ,{\rm age})$), we get:
for the galaxy C-14924 age$_{\rm old}=4.2^{+6.8}_{-3.4}$(95\%CL,max.likelihood) Gyr;
for the galaxy J-1010816 age$_{\rm old}=3.2^{+7.3}_{-2.5}$(95\%CL,max.likelihood) Gyr;
for the galaxy J-1050323 age$_{old}=9.7^{+1.3}_{-5.6}$ (95\% CL, max.likelihood) Gyr;
$<0.8$ Gyr (the age of the Universe at the corresponding redshift $z=6.90$) has a probability
$<4\times 10^{-7}$ ($>4.9\sigma $).

These results are almost identical to those without TP-AGB effect.
The fact that we have almost no variation of the results stems from the much higher relative error bars for our photometric points
with $\lambda _{\rm rest}>7000$ \AA \ with respect to the points with redder wavelengths, 
which contribute very little to the SED fit, and that only stars with $\sim 0.6-2.0$ Gyr
have important TP-AGB enhancement. 

\subsubsection{Galaxies older than the Universe?}

Globally, one or two cases among 31 galaxies older than the age of the Universe within 2$\sigma $ are 
statistically expected. However, the case of J-1050323 alone, older than the age of the Universe within $>4.7\sigma$, cannot be considered a statistical anomaly. Similar results were obtained by \citet{Mar25} for galaxies of similar redshift, though systematic errors were not evaluated in the same extension as we did here, and the sample of galaxies examined by \citet{Mar25} was
selected in a different way, containing very few LRDs while the objects analysed in our work are mostly LRDs.

\begin{figure}[h]
\vspace{0cm}
\centering
\includegraphics[width=8cm]{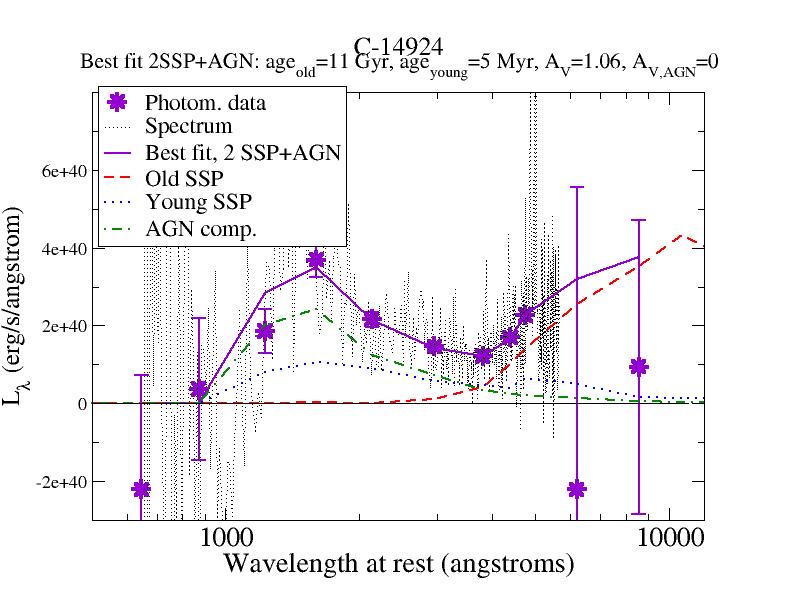}
\vspace{.2cm}
\includegraphics[width=8cm]{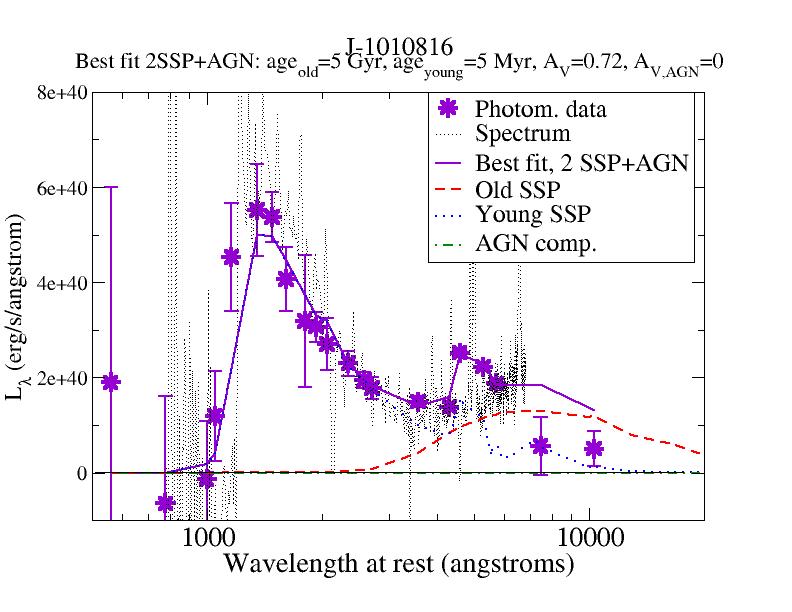}
\vspace{.2cm}
\includegraphics[width=8cm]{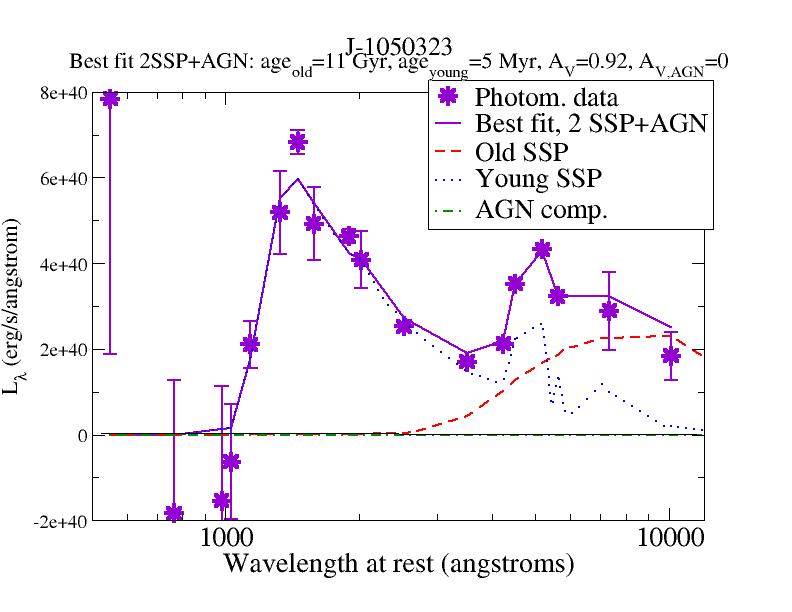}
\vspace{.2cm}
\caption{Best-fitting using two SSPs and one AGN 
of the SED at rest of the galaxies C-14924, J-1010816, and J-1050323 ($\chi ^2_{\rm red}=1.95$, 1.88, 5.82 
respectively). Spectra are available for the first two galaxies. Note that the age in these plots is the one with 'best fit' (lower $\chi ^2$); 
whereas the range indicated in the items of \S \ref{.old} are those corresponding to maximum likelihood statistics, with central value the one in which there is equal probability (50\%) to be higher or lower than it.}
\label{Fig:SEDgals}
\end{figure}

\section{Summary, discussion and conclusions}
\label{.disc}

Thirty-one {\it JWST} galaxies at $z=6-10$ were analysed with a large Balmer break, Ly-$\alpha $ break (V-shape; observable at near infrared bands) and very low emission at shorter wavelengths of far-UV at rest (corresponding to optical bands with {\it HST}), with the addition of mid-infrared fluxes when available, and spectra for 13 of them. In this type of galaxies, the 
SED fitting only with the photometric data
is useful to constrain the age of the galaxies (equal to the age of the oldest stellar population), given
that the degeneracy age-extinction is broken in the global analysis of both breaks (Paper I); 
spectral features alone, such as Balmer+4000 \AA \ break or CaII lines, may also provide some independent 
constraints of the 'average' age of the stellar populations.
We included metallicity as a free parameter in our SED fittings, but metallicity changes affect very 
little and cannot substitute the necessity of an old population.
The approach leaving the age of the oldest population as a free parameter without constraints serve both to determine the age of the galaxies
and to test the validity of $\Lambda $CDM cosmology. We may wonder how do the results and conclusions change if we limit the range of the age of the old population to the age of the Universe. The answer to this question is that the fits for some galaxies (especially the old galaxies of \S \ref{.old}, Fig. \ref{Fig:SEDgals}) would have a much worst fit (much larger $\chi^2$). These numerical experiments were also discussed in Paper I.

There are six main elements to take into account in the SED fitting: a dominant old stellar population, a young stellar population, emission lines associated with HII regions, AGN, interstellar extinction, intergalactic extinction due to neutral hydrogen in the reionization epoch. 
Extra stellar populations do not significantly improve the fits (\citet{Lop17}, Paper I). 
Paper I already did this kind of analysis, but here the analysis has been improved, 
introducing a larger number of galaxies of this type (31 instead of 13), and using {\it JWST} mid-infrared (5-26 $\mu $m) photometry and near-infrared spectra when available. Emission lines could be modelled using the information in the available spectra, and AGN with extinction $A_V>3$ is explored too.
Also, intergalactic extinction due to neutral hydrogen affecting wavelengths shorter than 1\,216 \AA -rest was here taken into account, whereas
it is not taken into account in Paper I.
As result of this more accurate analysis, ages are significantly lower than Paper I.

As discussed in Sect. \ref{.intro}, 
the presence of pronounced Balmer breaks make necessary
the old stellar component \citep{Set24,Ma25}, and this is what led to the results in this paper.
There are however other exotic ideas under discussion the recent literature claiming the possibility to produce
Balmer breaks in terms of AGNs surrounded by clumps of extremely dense gas \citep{Ina25}, 
or super-Eddington accretion AGNs \citep{Liu25}. This conflicts with our findings of a low AGN component,
and, moreover, SED of these V-shaped
galaxies always lie at the Balmer break, very unlikely to be produced in terms of  
AGN-only or AGN+young-stars models with a sum of blue continuum and red continuum
producing a minimum flux coincidentally at 3645 \AA -rest  \citep{Set24}.
Super-Eddington accretion is also excluded by X-ray observations \citep{Sac25}.
Anyway, here we have not explored these exotic possibilities in this paper.

From this work, it can be concluded that: 
\begin{enumerate}
\item A young stellar population (5-25 Myr) or AGN component is necessary to obtain the high observed relative fluxes around Ly-$\alpha $ break.
\item Within the stellar components (when AGN component is negligible), extinction in the young stellar population cannot be high, otherwise the Ly-$\alpha $ break could not 
be observed. A lower extinction for the younger population than for the old population would be unphysical, so the extinction of the old population should also be low ($A_V\lesssim 2$). This avoids the degeneracy age-dust at Balmer break and longer wavelengths. Further discussion at Paper I(Sect. 3.3). 
\item On average, a large AGN component ($A_3\gtrsim 1$), either with low or high extinction, 
can be rejected. 
The data allow to constraint and average $A_3$ (the ratio of AGN/old population fluxes at 5500 \AA $_{\rm rest}$) to be 
lower than 0.25 (95\% CL), much lower ratio
near the Balmer break.
Moreover, the line ratios are typical from starburst rather than AGN (Sect. \ref{.emlines}).
 Therefore, the hypothesis suggested by \citet{Bar24,Chw24,Li25} or W24 that these galaxies
are dominated by red AGNs, although possible for some individual cases, is not appropriate for the average of the 31 galaxies.

These objects are associated with the term 'little red dots' (LRDs): they are very red, they are very compact, at very high $z$.
And although some of these galaxies may be dominated by a massive AGN, spectroscopic analyses of this type of sources show a small ratio of AGN with a massive black hole
(only 2 of 13 have broad lines with $\sigma >1000$ km/s), in agreement with a 
previous analysis by \citet{Per24} who got 15\% of them  presenting spectral features of AGN.  
The nature of these
LRDs is not clear yet. 
The authors who have fitted the SED of these galaxies as a combination of stellar population and
red QSOs with similar ratios (e.g., W24, \citet{Ma25}) commonly adopt an assumption on extinction curves that
is not realistic.
Indeed, the remarkable observation that the inflection of the SED of these V-shaped
galaxies always lie at the Balmer break makes 
AGN-only or AGN+young-stars models unlikely given the data available \citep{Set24}; a sum of blue continuum and red continuum
should produce a minimum flux at any wavelength rather than always at 3645 \AA -rest.
Some interpretations of broad lines different from AGN were also proposed
(see Sect. \ref{.intro}).

\item Only a dominant old population ($\gtrsim $ 200 Myr) 
around Balmer break may explain its red color.
The age of these galaxies is analysed here. Tables 
\ref{Tab:age_sp} and \ref{Tab:age_vs_color} compiled the ages
calculated throughout this paper, either for the stacking of the 31 galaxies, 
or only with some selected colour $F277W-F444W$, either from the photometry
or from the best indicator with spectroscopy [$D_n(4000)$]. 
SED-photometry fitting gets better constraints for the minimum age, whereas
the maximum age is sometimes better limited by spectroscopy and other times by photometry.

Combining with weighted average the results of photometric (Table \ref{Tab:age_vs_color}, last column) 
and spectroscopic measurements (Table \ref{Tab:age_sp}), we get that
the average age of the galaxies of our sample is
\begin{equation}
0.61\pm 0.31(95\% CL)\ {\rm Gyr}
,\end{equation}
compatible to be younger than the average age of the $\Lambda $CDM universe $700^{+30}_{-20}$ Myr.
This corresponds to an estimated formation epoch $z_{\rm form.}>11.2$ (97.5\% CL).
For the galaxies with reddest color, F277W-F444W$>1.7$, the average age is
\begin{equation}
1.3^{+1.4}_{-0.8}(95\% CL)\ {\rm Gyr}
.\end{equation}
This is the average age, but the photometric age of some individual galaxies might be much larger.
Three of 31 galaxies present ages $\ge 600$ Myr (2$\sigma $): 4.2$^{+6.8}_{-3.4}$(95\%CL) Gyr,
2.5$^{+7.3}_{-1.9}$(95\%CL) Gyr and 9.5$^{+1.5}_{-5.6}$(95\%CL) Gyr respectively.
The last case (galaxy J-1050323) is incompatible to be younger than the corresponding age of the Universe
at its redshift $z=6.9$ ($t_{\rm \Lambda CDM, Univ}=0.8$ Gyr) at $>4.7\sigma $ level.

For the average redshift of these galaxies within $\Lambda $CDM, this would imply for the standard cosmological model that
these 31 galaxies and their first stellar populations were formed on average at $z_{\rm form.}>11.2$ (97.5\% CL).
\end{enumerate}

These numbers may be affected by some systematic errors associated to assumptions used here, though they
are expected to be small. 
Systematic errors may stem from:
\begin{itemize}
\item Only two stellar populations are considered. Numerical experiments in Paper I show that V-shaped SED fits are not improved
with a higher number of SSPs, thus an important systematic error in the age due to this assumption is unlikely.

\item We neglect the light contamination from other galaxies. This is justified because
the galaxies were selected to be isolated, so we do not expect an important systematic error due to contaminants.

\item TP-AGB continous emission enhancement at $\lambda _{\rm rest} >$ 7000 \AA \ is only relatively important 
for ages larger than 0.6-1.0 Gyr \citep{Mar05,Lop17,Mar25,Lu25}. Our galaxies in the standard cosmological model should be younger than the age of the universe, $\approx 0.7$ Gyr, (and most likely younger than $\approx 0.5$ Gyr, given the formation time needed for these massive galaxies), so no important TP-AGB features should be observed in this case. 
Only if the cosmological model is different from the
standard $\Lambda $CDM  and some of the galaxies (like J-1050323) are older than 0.6 Gyr would the TP-AGB potentially
affect the wavelengths redder than 7\,000 \AA , which should then be investigated. In any case,
the much higher relative error bars for our photometric points with $\lambda _{\rm rest}>7000$ \AA \ 
with respect to the points with redder wavelengths makes the TP-AGB effect very small in the global SED fit.
In the case of the three oldest galaxies, we found negligible the effect of TP-AGB enhancement (Sect. \ref{.TP-AGB}).
Although we have spectra of some galaxies, their resolution and signal/noise are not enough
to measure absorption lines, so we cannot investigate typical TP-AGB spectral features \citep{Lu25}.
Said in other words, we cannot claim that TP-AGB effects could be identified or evaluated given the current data quality.

\item We assume an IMF of Chabrier adopted by GALAXEV model \citep{Bru03}. Systematic errors due to the variation of IMF are not important either \citep{Mar25}. 

\item Apart from IMF and TP-AGB modelling, other differences in different synthesis of stellar population models may give
different ages, though within the same order of magnitude (e,g,, \citet{Lab23,Bar24}; Paper I).  

\item Emission lines are assumed to be like in the average spectrum (Sect. \ref{.emlines}). This is a reasonable approximation for the average stacked photometry, but it may introduce
a significant variation of the age for individual galaxies when the lines are strong enough to contribute with an important ratio of flux in the photometry wide-intermediate band filters. We have carried out new SED fits for the candidate galaxy of oldest galaxy, J-1050323 (Sect. \ref{.oldemlines}), and we see the effect of changing the
relative amplitude of emission lines can slight change the range of ages, but it still needs to be much older than the age of the Universe.  

\end{itemize}
Nonetheless, other systematic errors may arise under some changes in the assumptions of our analysis.
The results obtained in this paper indicate a certain tension between the ages of some galaxies 
and the age of the Universe at their corresponding redshifts. This is not a definitive result,
but something that should warrant further research. 
Larger samples of objects as well as improvements in stellar population models, and abundance, properties and 
distribution of the dust may help to solve these discrepancies.
\\

{\bf Acknowledgements:}
Thanks are given to the anonymous referee and the editor Claudia Maraston for helpful comments.
We acknowledge support from the Spanish Ministerio de
Ciencia, Innovación y Universidades (MICINN) under grant numbers PID2021-129031NB-I00 and PID2022-141915NB-C21.
This work is based on observations made with the NASA/ESA/CSA James Webb Space Telescope. The data were obtained from the Mikulski Archive for Space Telescopes at the Space Telescope Science Institute, which is operated by the Association of Universities for Research in Astronomy, Inc., under NASA contract NAS 5-03127 for {\it JWST}. 
This work made use of data from the Cosmic Evolution Early Release Science Survey (CEERS) and 
{\it JWST} Advanced Deep Extragalactic Survey (JADES).
\\

DATA AVAILABILITY: The data underlying this article are available in the article (Table \ref{Tab:newfluxes}) for Mid Infrared Data; 
in \citet{Lab23} and in its online supplementary material
for CEERS data; in https://archive.stsci.edu/hlsp/jades for JADES data; and https://dawn-cph.github.io/dja/spectroscopy/nirspec/ for spectra.
\\

\section*{ORCID for authors}

M. L\'opez-Corredoira: 0000-0001-6128-6274

C. M. Guti\'errez: https://orcid.org/0000-0001-7854-783X

\end{document}